\documentclass[journal,draftclsnofoot,onecolumn,12pt]{IEEEtran}
\usepackage{amsthm,amssymb,graphicx,multirow,amsmath,color,amsfonts}
\usepackage[update,prepend]{epstopdf}
\usepackage[noadjust]{cite}
\usepackage[latin1,utf8]{inputenc}
\usepackage{tikz}
\usepackage{bbm} 
\usepackage{pdfpages}
\usepackage{tabulary}
\usepackage{multirow}
\usepackage{comment}
\usepackage{subfigure}
\usepackage{float}
\DeclareUnicodeCharacter{0177}{\^y}

\def\nb0{{\mathbf{0}}}
\def\nb1{{\mathbf{1}}}

\newtheorem{lemma}{Lemma}

\newtheorem{definition}{Definition}

\newtheorem{theorem}{Theorem}

\newtheorem{cor}{Corollary}

\newtheorem{remark}{Remark}

\allowdisplaybreaks 
\begin{document}
\title{On the Downlink SINR Meta Distribution of UAV-assisted Wireless Networks}
\author{
	Yujie Qin, Mustafa A. Kishk, {\em Member, IEEE}, and Mohamed-Slim Alouini, {\em Fellow, IEEE}
	\thanks{Yujie Qin and Mohamed-Slim Alouini are with Computer, Electrical and Mathematical Sciences and Engineering (CEMSE) Division, King Abdullah University of Science and Technology (KAUST), Thuwal, 23955-6900, Saudi Arabia
		Arabia. Mustafa Kishk   is with the Department of Electronic Engineering, Maynooth University, Maynooth, W23 F2H6, Ireland. (e-mail: yujie.qin@kaust.edu.sa; mustafa.kishk@mu.ie; slim.alouini@kaust.edu.sa).} 
	
}
\date{\today}
\maketitle
\begin{abstract}
	The meta distribution of the signal-to-interference-plus-noise ratio (SINR) provides fine-grained information about each link's performance in a wireless system and the reliability of the whole network. While the UAV-enabled network has been studied extensively, most of the works focus on the spatial average performance, such as coverage probability, while SINR meta distribution has received less attention.  In this paper, we use the SINR meta distribution for the first time to systematically analyze the improvement and the influence of deploying UAVs on the reliability of a wireless network. We first derive the $b$-th moments of the conditional success probability of the UAV-enabled network and give the approximated expressions derived by Gil-Pelaez theorem  and the beta approximation of the meta distribution. Our numerical results show that deploying UAVs in wireless networks in most cases can greatly improve the system reliability, which denotes the fraction of users achieving cellular coverage, especially for the spatially-clustered users. In addition, establishing LoS links is not always beneficial since it also increases the interference. For instance, with the increase of the SINR threshold, the system reliability of a high LoS probability environment decreases dramatically and it is even lower than a low LoS probability environment.  We also show that in highrise urban areas, UAVs can help in establishing extremely reliable (very high SINR) links. 
\end{abstract}
\section{Introduction}
As a result of recent technology advancements and cost reduction, unmanned aerial vehicles (UAVs, or so-called drones) are expected to play an essential role in potentially improve the performance of the next-generation wireless networks \cite{sekander2018multi,mozaffari2019tutorial,8579209,7470933}.
The primary benefits of UAVs include that they can efficiently function as aerial base stations (BSs) with high relocation flexibility based on dynamic traffic demands, improve the coverage probability by establishing line-of-sight (LoS) links with ground users \cite{8675384,8833522}. Moreover, drones can be deployed in dangerous environments or in natural disasters, where the traditional terrestrial BSs (TBSs) are more likely to be unavailable or heavily damaged \cite{matracia2021coverage}, while drones can provide stable and flexible connectivity, which makes them a feasible and practical alternative. Moreover, at places where the spatial distribution of active users continuously change with time, the dynamic deployment of UAVs is more efficient compared to terrestrial communications with static BSs since they have the capability to optimize their locations in real-time \cite{9500746}. Under the circumstance in which the users' locations exhibit a certain degree of clustering \cite{afshang2017nearest}, UAVs can assist TBSs to deliver user clusters with reliable network coverage and complement existing cellular networks by providing additional capacity \cite{9205314}. 

Generally, one of the most popular tools of analyzing UAV-assisted networks is stochastic geometry, which has been widely utilized in modeling, characterizing, and obtaining design insights of the wireless networks with randomly placed nodes \cite{lu2021stochastic,elsawy2017meta}. 
For analyzing UAV-involved networks, most of the stochastic geometry-based analyses are confined to the spatial average, such as coverage probability \cite{elsawy2016modeling,haenggi2021meta1,haenggi2021meta2}, while the performance from the perspective of each user is ignored. For instance, the conditional success probability of an individual user, which denotes the probability of the user being successfully served by the BS conditioned on the realization, which provides information about end-to-end delay and quality-of-service, is also very important.  Besides, deploying UAVs in wireless cellular networks cannot always improve the system performance due to the increasing of interference. For example, deploying UAVs in wireless networks with different SINR threshold values, environments (e.g., highrise, urban, dense, and suburban), and user distributions has different impacts.
To this end,  it is crucial to obtain key information about the "distribution of success probability of the individual link in a given network" \cite{8648502}, which reveals the reliability and quality of service (QoS) of the network and is a fundamental design objective for cellular operators. This new and fundamental performance metric is called the SINR meta distribution \cite{haenggi2015meta}, defined as the complementary cumulative distribution function (CCDF) of success probability conditioned on the point process realization.

In this work, we are interested in analyzing the impact of the deployment of UAVs in cellular networks on the reliability of UAV-assisted networks. We use stochastic geometry tools to model the locations of UAVs, TBSs, and two types of users: (i) Mat\'{e}rn cluster process (MCP) and (ii) Poisson point process (PPP). By computing the new and fundamental performance metric: the SINR meta distribution, we obtain the system insights from the perspective of each user and the improvement of the system's reliability by deploying UAVs.

\subsection{Related Work}

Literature related to this work can be categorized into: (i)  stochastic geometry-based frameworks for UAV wireless networks and (ii) the SINR meta distribution-related analysis. A brief discussion on related works in each of these categories is discussed in the following lines.

{\em Stochastic Geometry-based Literature.} Stochastic geometry is a strong mathematical tool that enables characterizing the statistics of various large-scale wireless networks. The authors in \cite{elsawy2016modeling,elsawy2013stochastic} presented a  tutorial on the fundamental concepts of the point process, modeling the interference in large-scale networks, and a comprehensive survey on single-tier, multi-tier, and cognitive cellular networks. Authors in~\cite{8833522} studied a heterogeneous network composed of users, terrestrial and aerial BSs spatially distributed according to three independent Poisson point processes (PPPs). Under their system setup, the authors accurately characterized and derived the Laplace transform of the interference from aerial and terrestrial BSs, downlink coverage probability, and average data rate. Authors in \cite{9153823,9444343,9773146} modeled the locations of UAVs and charging stations by two independent PPPs and modified the definition of coverage probability based on queuing theory by considering the energy limitation of UAVs. Authors in \cite{sekander2020statistical} considered renewable energy-powered UAVs, which can harvest energy from solar or wind resources, and derived the probability density function (PDF) and cumulative density function (CDF) of harvested energy as well as the outage probability. Besides harvesting energy, authors in \cite{kishk20203,9205314,lou2021green} studied the tethered UAV, which is physically connected to a ground station. While the tether provides the UAV with a stable power supply and reliable data rate, it highly restricts the mobility and freedom of UAVs. Besides, the authors in \cite{9456851} provided a comprehensive overview  of the latest research efforts on integrating UAVs into cellular networks and highlighted important directions for further investigation in future work. 

{\em Meta Distribution-based Literature.} Authors in \cite{haenggi2015meta} used the SIR meta distribution to study the downlink performance of Poisson bipolar networks with ALOHA channel access. They first derived the conditional success probability and then gave both the exact expression of the meta distribution and a highly accurate approximation, named beta approximation. The concept of the conditional success probability was first given in \cite{baccelli2010new}, which was used in computing the local delay. 
 In \cite{haenggi2021meta1}, the author gave the detailed definition of SIR  meta distribution and some examples of computing SIR meta distribution, such as Poisson bipolar networks and PPP networks. 
 Besides, a closed-form of SIR meta distribution was provided in  \cite{haenggi2021meta2} which only considered the nearest interference in Poisson and Poisson bipolar networks. 
 The meta distribution of downlink SIR in a Poisson cluster process-based heterogeneous networks (HetNet) was analyzed in \cite{saha2020meta}, in which the authors considered a $K$-tier PPP network. Interestingly, they used mapping theorem \cite{madhusudhanan2016analysis,stoyan2013stochastic} to map the interference from $i$-th tiers onto one tier, which forms a new and unknown distribution. The authors in  \cite{elsawy2017meta} studied the meta distribution in uplink case, where the locations of BSs are modeled by PPP and the transmission is with fractional path-loss inversion power control. In \cite{mankar2019meta}, the authors characterized the meta distribution of the downlink SIR for the typical cell in the case that the BSs are modeled by PPP. For a general cellular network, simple approximations of the SIR meta distribution were analyzed in \cite{kalamkar2019simple}. The authors showed that the meta distribution of a general network can be obtained by a shift of a PPP network, where the shift is a function of the mean interference-to-signal ratio. The SIR meta distribution   in the case that base station cooperation was analyzed in \cite{cui2017sir}, in which the authors showed the benefits of different cooperation schemes and the impact of the number of cooperating base stations. SIR meta distribution of $K$-tier downlink heterogeneous cellular networks (HCNs) with cell range expansion was analyzed in \cite{wang2018sir} and cellular networks with power control were studied in \cite{wang2017meta}. Authors in \cite{feng2020separability} show that the separable form is a good approximation of SIR meta distribution in Ginibre and triangular lattice networks when the SINR threshold is very large. Authors in \cite{9170570} studied the SIR/SNR meta distribution in a cellular network with coexisting sub-6GHz and millimeter wave spectrums.

 Different from the existing literature, which mainly focus on the spatial average performance of UAV-enabled networks, this paper studies the SINR meta distribution of downlink transmission of UAV-enabled networks. We systematically investigate the reliability of UAV-enabled networks by using the SINR meta distribution for the first time. Motivated by the fact that future wireless networks are expected to be a mix of TBSs and UAVs, we analyze the reliability of the coexistence of UAVs and TBSs networks. 

\subsection{Contribution}
Unlike most of the work on UAV-assisted networks, which only consider a spatial average performance, we develop a stochastic geometry-based framework to analyze the downlink performance from the perspective of each link for the first time. We compare the UAV-enabled network with the TBS-only network and show how deploying UAVs affects system reliability.
The contributions of this paper are:

\begin{itemize}
	\item Considering two types of the most widely used user distributions, the Poisson point process (PPP) and Mat\'{e}rn cluster process (MCP), we give the expressions for the moments $M_b(\theta)$ of UAV and TBS-mixed networks, in which the air-to-ground channels are modeled by Nakagami-m fading with different LoS/NLoS scale and shape parameters, and ground-to-ground channels are modeled by Rayleigh fading, where $\theta$ is SINR threshold. 
	
	\item We provide the theoretical equations for the exact meta distributions of the UAV and TBS-mixed networks in the case of two types of user distributions. Considering the high complexity of computing the exact meta distribution, we use the commonly used beta approximation and show that the beta approximation is highly accurate in UAVs' networks for a large range of $\theta$.
	
	\item By comparing with the TBS-only networks, we theoretically prove that the deployment of UAVs in wireless cellular networks can dramatically improve the system's reliability in some scenarios. However, establishing LoS links does not always benefit the users. For instance, in the case of a highrise area, deploying a UAV for each cluster seems only slightly improving the reliability. However, once a LoS link established between the user and the cluster UAV, this channel can achieve an extremely high SINR. These results can be beneficial in Internet of Things (IoT) networks, which highly require reliable data transmission. In the case of urban areas,  operators need to decrease UAVs' altitude and density to maintain a reliable network.
		
	\item  Besides the impact of environment on the system's reliability, we provided both the simulation and analysis results for the meta distribution of two types of user distributions, spatially-clustered or PPP distributed. We showed that the system benefit more if the users are spatially-clustered. Besides, we also showed the impact of altitudes and densities of UAVs on the system's reliability. 
\end{itemize}

We would like to clarify that, while the propagation model, point process models, and beta approximation are well-developed, the main contributions of this work are theoretically analyzing the influence of UAVs on the reliability of wireless networks from the perspective of each link, and applying the concept of SINR meta distribution in UAV-assisted networks for the first time.


\section{System Model}
We consider a wireless cellular network comprising of UAVs and TBSs, and focus on the analysis of the reliability of the downlink performance. Generally, MCP and PPP are two most widely useful point processes in modeling the user distributions \cite{saha2017enriched} and the motivation of modeling the locations of users by a MCP can be found in \cite{afshang2017nearest,ganti2009interference,afshang2016nearest}. In this work, we consider two types of user distributions: (i) MCP, where the users are uniformly distributed within the user clusters and the centers of each cluster are spatially distributed according to a PPP, (ii) PPP, where the users are randomly located within the whole plane. 

\begin{figure}[ht]
	\includegraphics[width=1\columnwidth]{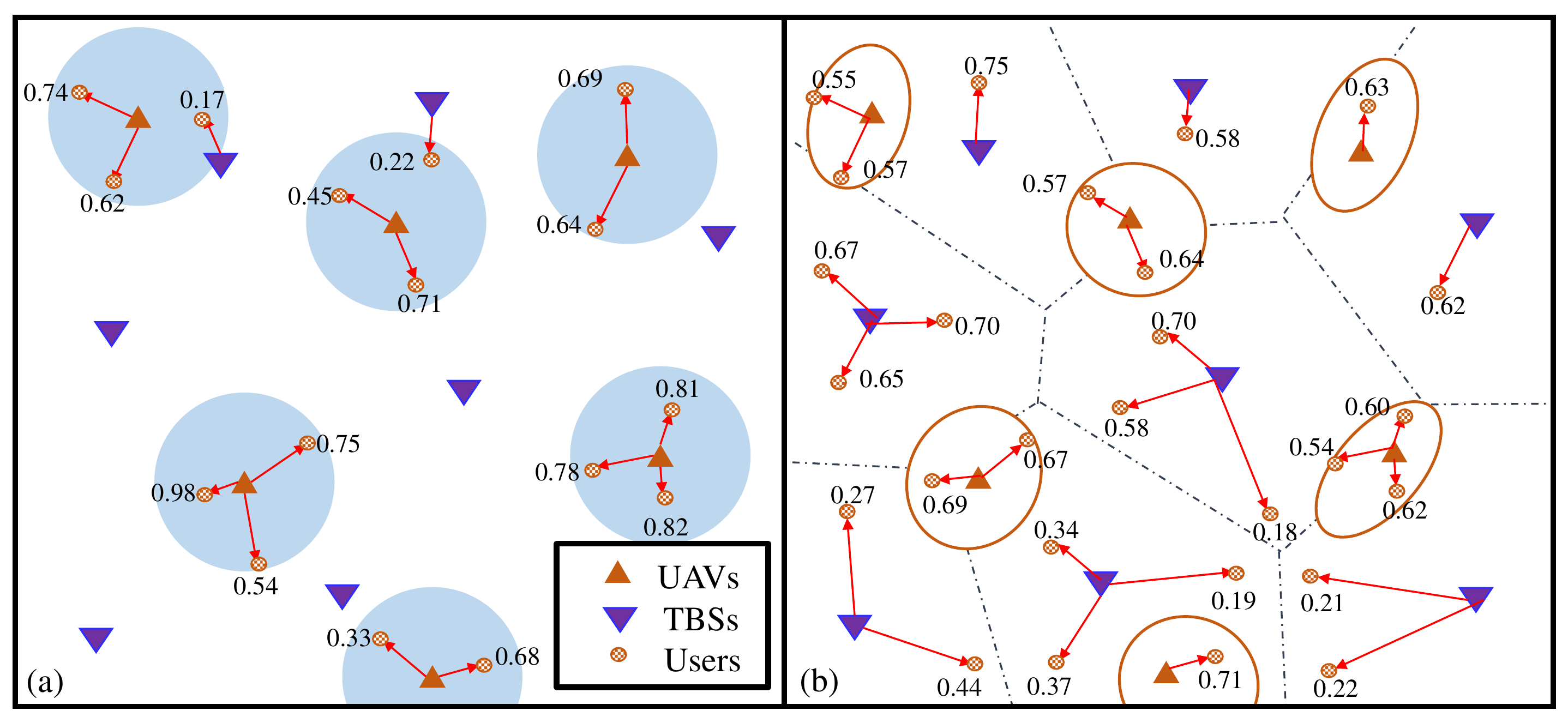}
	\caption{Illustration of the system models: realizations of Poisson cellular networks with user distributed according to \textbf{(a)} MCP  and \textbf{(b)} PPP. The communication channels are indicated by arrows, UAVs/TBSs are by upper/lower triangles and users are denoted by markers, respectively. The number next to each arrow is its success probability (averaged over fading). }
\end{figure}

\begin{definition}[System Model 1: MCP Users]
 In the case of users are spatially clustered (the locations of users are modeled by a MCP), we consider that the locations of user cluster centers are modeled by PPP $\Phi_{u,1}$ with density $\lambda_{u,1}$, and the clusters are modeled as disks with radii $r_c$.  UAVs are hovering at a fixed altitude $h$ above each cluster center to provide service. Note that since each user cluster center has a UAV deployed, the density of UAVs is $\lambda_{u,1}$ and the locations of UAVs are denoted by $\Phi_{u,1}$.
\end{definition}
\begin{definition}[System Model 2: PPP Users]
	In the case of users are PPP distributed, we consider that the locations of users and UAVs are modeled by two independent  PPPs. Let  $\Phi_{u,2}$ be the point set of the locations of UAVs with density $\lambda_{u,2}$.
\end{definition}
In the above two system models, the locations of TBSs are modeled by another independent PPP $\Phi_t$ with density $\lambda_t$ and let $\Phi_{\rm user}$ be the point set that contains the locations of users. 

\subsection{User Association}
  We assume that in the case of system model 1, the user associates with the cluster UAV, the UAV deployed at the user's cluster center, or the nearest TBS that provides the strongest average received power. Note that the association policy of system model 1 is known as closed access \cite{dhillon2012modeling}, in which each UAV only serves the users in its own cluster. In the case of system model 2, user associates with the UAV or TBS that provides the strongest average received power. 
Let $R_u$ be the Euclidean distance between the user and the cluster UAV in the first system model, and $R_t$ is the distance between the user and the nearest TBS in both system models.

When the reference user associates with the UAV, the received power is given by,
\begin{align}\label{eq_1}
	p_{ u} = \left\{
	\begin{aligned}
		p_{u, l} &= \eta_{  l}\rho_{  u}G_{  l}R_{u}^{-\alpha_{ l}}, \text{in the case of LoS},\\
		p_{u, n} &= \eta_{ n}\rho_{ u}G_{ n}R_{u}^{-\alpha_{ n}}, \text{in the case of NLoS},\\
	\end{aligned}
	\right.
\end{align}
in which $\rho_{ u}$ is the transmit power of UAVs, $\alpha_{ l}$ and $\alpha_{ n}$ are path-loss exponent, $G_{ l}$ and $G_{ n}$ present the fading gains that follow gamma distribution with shape and scale parameters ($m_{ l},\frac{1}{m_{ l}}$) and ($m_{ n},\frac{1}{m_{ n}}$), $\eta_{ l}$ and $\eta_{ n}$ denote the mean additional losses for LoS and NLoS transmissions, respectively. The occurrence probability of a LoS link established by the reference user and the UAV with a horizontal distance $r$ is given in ~\cite{al2014optimal} as
\begin{align}
	P_{ l}(r) & =  \frac{1}{1+e_1 \exp(-e_2(\frac{180}{\pi}\arctan(\frac{h}{r})-e_1))},\label{pl_pn}
\end{align}
where $e_1$ and $e_2$ are two environment variables (e.g., suburban, urban, dense urban, and highrise urban), and $h$ is the altitude of the UAV. Consequently, the probability of NLoS link is $P_{ n}(r)=1-P_{  l}(r)$.

When the user  associates with the nearest TBS, the received power is
\begin{align}
	p_{ t} &= \rho_{ t} H R_{ t}^{-\alpha_{ t}},
\end{align}
where $H$ is the fading gain that follows exponential distribution with average of unity, and $\rho_{ t}$ is the transmit power of TBSs.

\subsection{SINR Meta Distribution}
We randomly select the location of one user $u_o\in \Phi_{\rm user}$ as our reference user. Conditioned on the realizations of $\Phi_{  t}$, $\Phi_{u,\{1,2\}}$ and $\Phi_{\rm user}$, the conditional success probability of the reference link established by reference user and its serving BS, is given by
\begin{align}
	P_s(\theta) &= \mathbb{P}({\rm SINR}>\theta|\Phi_{ t},\Phi_{u,\{1,2\}},\Phi_{\rm user}),
\end{align}
where ${\rm SINR}$ is signal to interference plus noise ratio. Since we have two system models and the interference terms are slightly different for each system, we give the expressions for SINR term in detail  later in Section \ref{sec_PerAna}.

\begin{definition}[$b$-th Moments]Consequently, by taking the expectation over all the random variables (locations and distances) the $b$-th moment of the conditional success probability is defined as
	\begin{align}
		M_b(\theta) = \mathbb{E}[P_{s}^{b}(\theta)].
	\end{align}
\end{definition}

For arbitrary realizations of $\Phi_{\rm user}$, $\Phi_{u,\{1,2\}}$ and $\Phi_b$, we would like to analyze the percentile of the links exceeding a predefined SINR threshold $\theta$, which is defined as the SINR meta distribution. With that being said, our goal is to obtain the percentiles of users that achieve downlink coverage (SINR above $\theta$) in an arbitrary but fixed realization of networks.   Note that the randomness in $P_s(\theta)$ comes from the channel fading. 
\begin{definition}[Meta Distribution]
	The SINR meta distribution, which is the CCDF of conditional success probability, of downlink is defined as 
	\begin{align}
		\bar{F}_{P_s}(\theta,\gamma) = \mathbb{P}(P_s(\theta) > \gamma),\label{eq_MetaF}
	\end{align}
	where $\gamma \in [0,1]$.
\end{definition}

\section{Performance Analysis}
\label{sec_PerAna}
The goal of this work is to analyze the improvement of system performance by deploying UAVs in  wireless cellular networks. For each of the system models, we derive the SINR meta distributions and then compared it with (i) TBS-only networks, and (ii) scenario 1, where all the aerial BSs (UAVs) are converted into ground stations. Scenario 1's motivation is to investigate the benefit of the altitude of the deployed UAVs, compared to deploying terrestrial BSs at the same locations. More details are provided in the following text.

\subsection{MCP Distributed Users}
In this section, we provide the analysis for the system model 1: the locations of users are modeled by MCP, and UAVs are deployed at a fixed altitude $h$ above the cluster centers to provide service. To do so, we first need to derive the distance distributions which are shown in the following lemma.

\begin{lemma}[Distance Distribution] The probability density function of the distances between the reference user and the UAV in its cluster center, the nearest TBS, denoted by $f_{ R_u}(r)$ and $f_{ R_t}(r)$, respectively, are given by
	\begin{align}
		f_{ R_{u}}(r) &= \frac{2r}{r_{c}^2}, \quad h\leq r \leq \sqrt{r_c^2+h^2},\nonumber\\
		f_{ R_t}(r) &= 2\pi r\lambda_t\exp(-\pi\lambda_t r^2).
	\end{align}
Thus, the CDF of $R_t$ is,
\begin{align}
	F_{ R_t}(r) &= 1-\exp(-\pi r^2 \lambda_t).
\end{align}
\end{lemma}
Conditioned on the serving BS, we need to know the distance to the nearest interfering BSs, which is given in the following lemma.
\begin{lemma}[Nearest Interfering BSs] Given that the reference user is associated with the cluster UAV at a distance $R_u$ away, in the case of LoS/NLoS link, the nearest interfering TBS is at least $	d_{lt}(R_u)$ and $d_{nt}(R_u)$ away, respectively, given by
	\begin{align}
		d_{lt}(R_u) &= (\frac{\rho_t}{\rho_u\eta_l})^{\frac{1}{\alpha_t}}R_{u}^{\frac{\alpha_l}{\alpha_t}},\nonumber\\
		d_{nt}(R_u) &= (\frac{\rho_t}{\rho_u\eta_n})^{\frac{1}{\alpha_t}}R_{u}^{\frac{\alpha_n}{\alpha_t}}.
	\end{align}

\end{lemma}
\begin{IEEEproof}
Above equations follow from solving the equations of average received power: $\mathbb{E}[p_t]>\mathbb{E}[p_u]$, which is averaging over fading while conditioned on $R_u$.	
\end{IEEEproof}
\begin{remark}
	 Note that $d_{lt}(R_u)$ and $d_{nt}(R_u)$ are useful in identifying the user association events. For example, when the cluster UAV is NLoS and the distance is $R_u$, if the distance to the nearest TBS $R_t$ is greater than  $d_{nt}(R_u)$, the user associates with cluster NLoS UAV. Otherwise, the user associates with TBS. 
\end{remark}

Recall that in the system model 1, the user associates with the cluster UAV or the nearest TBS. Hence, the SINR at the reference user is
\begin{align}
	{\rm SINR} = \frac{\max{(p_u,p_t)}}{I+\sigma^2},
\end{align}
where $\sigma^2$ is the noise power and $I$ is the aggregate interference. In this system model, the user can (i) associate with cluster UAV in the case that it is LoS, (ii) associate with cluster UAV in the case that it is NLoS, (iii) associate with the nearest TBS in the case that cluster UAV is LoS, and (iv) associate with the nearest TBS in the case that cluster UAV is NLoS. Therefore, the conditional success probability can be written as
\begin{align}
	&P_s(\theta)= \mathbb{P}({\rm SINR}>\theta|\Phi_{  t},\Phi_{u,1},\Phi_{\rm user})= \mathbb{P}\bigg(\frac{\max{(p_u,p_t)}}{I+\sigma^2}>\theta|\Phi_{  t},\Phi_{u,1},\Phi_{\rm user}\bigg)  \nonumber\\
	&= \mathbbm{1}({\rm LoS})\mathbbm{1}(R_t>d_{lt}(R_u))\mathbb{P}\bigg(\frac{p_{u,l}}{I_{u,l}+\sigma^2}>\theta\bigg)+\mathbbm{1}({\rm NLoS})\mathbbm{1}(R_t>d_{nt}(R_u))\mathbb{P}\bigg(\frac{p_{u,n}}{I_{u,n}+\sigma^2}>\theta\bigg)\nonumber\\
	&\quad+\mathbbm{1}({\rm LoS})\mathbbm{1}(R_t<d_{lt}(R_u))\mathbb{P}\bigg(\frac{p_t}{I_{t,l}+\sigma^2}>\theta\bigg)+\mathbbm{1}({\rm NLoS})\mathbbm{1}(R_t<d_{nt}(R_u))\mathbb{P}\bigg(\frac{p_t}{I_{t,n}+\sigma^2}>\theta\bigg)\nonumber\\
	&=P_{s,l}(\theta)+P_{s,n}(\theta)+P_{s,tl}(\theta)+P_{s,tn}(\theta),\label{eq_ps_sys1}
\end{align}
 where $\mathbbm{1}({\rm LoS})$ and $\mathbbm{1}({\rm NLoS})$ are indicator functions which denote the events that cluster UAVs are LoS/NLoS, respectively, and $\mathbbm{1}(R_t>d_{nt}(R_u))$ and $\mathbbm{1}(R_t>d_{lt}(R_u))$ are functions of distances, hence, indicate the association events of users. For example, $\mathbbm{1}({\rm LoS})\mathbbm{1}(R_t>d_{lt}(R_u))$ indicates that the cluster UAV is LoS and the average received power from the nearest TBS is lower than cluster LoS UAV. Therefore, the user associates with the cluster LoS UAV.   Let $u_0$ be location of the cluster UAV from the perspective of the reference user, the aggregate interference terms are
\begin{align}
	I_{u,\{l,n\}} =& \sum_{N_i\in\Phi_{ u_n,1}\setminus \{u_0\}}\eta_{n}\rho_{u}G_{n,N_i}D_{  N_i}^{-\alpha_{ n}}+\sum_{L_j\in\Phi_{u_l,1}\setminus \{u_0\}}\eta_{l}\rho_{u}G_{l,L_j}D_{ L_j}^{-\alpha_{ l}}+\sum_{T_k\in\Phi_{ t}}\rho_{t}H_{T_k}D_{ T_k}^{-\alpha_{t}},\nonumber\\
	I_{t,l} =& \sum_{N_i\in\Phi_{ u_n,1}}\eta_{ n}\rho_{ u}G_{ n,N_i}D_{ N_i}^{-\alpha_{ n}}+\sum_{L_j\in\Phi_{ u_l,1}}\eta_{l}\rho_{u}G_{l, L_j}D_{  L_j}^{-\alpha_{ l}}+\sum_{T_k\in\Phi_{  t}\setminus \{t_0\}}\rho_{ t}H_{T_k}D_{ T_k}^{-\alpha_{ t}} + \eta_{ l}\rho_{ u}G_{ l}R_{ u}^{-\alpha_{ l}},\nonumber\\
	I_{t,n} =& \sum_{N_i\in\Phi_{ u_n,1}}\eta_{ n}\rho_{ u}G_{ n,N_i}D_{ N_i}^{-\alpha_{ n}}+\sum_{L_j\in\Phi_{ u_l,1}}\eta_{ l}\rho_{ u}G_{ l, L_j}D_{ L_j}^{-\alpha_{ l}}+\sum_{T_k\in\Phi_{ t}\setminus \{t_0\}}\rho_{ t}H_{T_k}D_{ T_k}^{-\alpha_{ t}} +
	\eta_{ n}\rho_{ u}G_{ n}R_{ u}^{-\alpha_{ n}},\nonumber
\end{align}
where $G_l$ and $G_n$ are described in (\ref{eq_1}),  $D_{\{\cdot\}}$ denotes the Euclidean distances between the interfering BSs to the reference user and $G_{l,\{\cdot\}}$, $G_{n,\{\cdot\}}$ and $H_{\{\cdot\}}$ denote the fading gains from different interferers. $\Phi_{ u_l,1}$ and  $\Phi_{ u_n,1}$ are subsets of $\Phi_{u,1}$ which present the locations of the interfering LoS UAVs and NLoS UAVs, respectively. Note that for different users, the subsets $\Phi_{ u_l,1}$ and  $\Phi_{ u_n,1}$ are different but $\Phi_{u,1}$ keeps the same. $t_0$ denotes the location of the serving (nearest) TBS and the difference between $I_{u,l}$ and $I_{u,n}$ is the distance to the nearest interfering TBS, $d_{nt}(R_u)$ and $d_{lt}(R_u)$, respectively.

\begin{remark}\label{rem_indicator}
The indicator functions in (\ref{eq_ps_sys1}) are widely replaced by the association probability in literature, by simply taking the expectation over the distance and writing as a function of CCDF: $\mathbbm{1}(LoS)\mathbbm{1}(R_t>d_{lt}(R_u)) = P_l(R_u)\mathcal{A}_{LoS}(d_{lt}(R_u))$, where $\mathcal{A}_{LoS}(d_{lt}(R_u)) = 1-F_{ R_t}(d_{lt}(R_u)) = \bar{F}_{ R_t}(d_{lt}(R_u))$. However, we here must write it as indicator functions because we need to compute the $b$-th moment of $P_s(\theta)$. In the case of indicator functions, we can avoid computing the cross terms which are complex and not accurate. More details are provided later in Remark \ref{rem_bth_moment}.
\end{remark}
\begin{lemma}[Conditional Success Probability] The conditional success probability of the reference link is given by
	\begin{align}
		P_{s,l}(\theta)&\approx 	\sum_{k_l=1}^{m_l}\binom{m_l}{k_l}(-1)^{k_l+1}\mathbbm{1}({\rm LoS})\mathbbm{1}(R_t>d_{lt}(R_u))\prod_{T_k\in\Phi_{ t}}f_{1}(s_2(l,k_l,R_u) ,D_{T_k},\alpha_t)\nonumber\\
		&\prod_{e\in\{l,n\}}\prod_{u_i\in\Phi_{ u_e,1}\setminus \{u_0\}}f_{2}(m_e,s_2(l,k_l,R_u),\eta_e,D_{u_i},\alpha_e)\exp(-s_2(l,k_l,R_u)\sigma^2) ,\label{eq_ps_l}\\
		P_{s,n}(\theta)&\approx 	\sum_{k_n=1}^{m_n}\binom{m_n}{k_n}(-1)^{k_n+1}\mathbbm{1}({\rm NLoS})\mathbbm{1}(R_t>d_{nt}(R_u))\prod_{T_k\in\Phi_{ t}}f_{1}(s_2(n,k_n,R_u) ,D_{T_k},\alpha_t)\nonumber\\
		&\prod_{e\in\{l,n\}}\prod_{u_i\in\Phi_{ u_e,1}\setminus \{u_0\}}f_{2}(m_e,s_2(n,k_n,R_u),\eta_e,D_{u_i},\alpha_e)\exp(-s_2(n,k_n,R_u)\sigma^2), \label{eq_ps_n}\\
		P_{s,tl}(\theta)&\approx \mathbbm{1}({\rm LoS})\mathbbm{1}(R_t<d_{lt}(R_u))f_{2}(m_l,s_t(R_u),\eta_l,R_u,\alpha_l)\prod_{T_k\in\Phi_{ t}\setminus \{t_0\}}f_{1}(s_t(R_u) ,D_{T_k},\alpha_t)\nonumber\\
		&\prod_{e\in\{l,n\}}\prod_{u_i\in\Phi_{ u_e,1}}f_{2}(m_e,s_t(R_u),\eta_e,D_{u_i},\alpha_e)\exp(-s_t(R_u)\sigma^2),\label{eq_ps_tl}\\
		P_{s,tn}(\theta)&\approx 	\mathbbm{1}({\rm NLoS})\mathbbm{1}(R_t<d_{nt}(R_u))f_{2}(m_n,s_t(R_u),\eta_n,R_u,\alpha_n)\prod_{T_k\in\Phi_{ t}\setminus \{t_0\}}f_{1}(s_t(R_u) ,D_{T_k},\alpha_t)\nonumber\\
		&\prod_{e\in\{l,n\}}\prod_{u_i\in\Phi_{ u_e,1}}f_{2}(m_e,s_t(R_u),\eta_e,D_{u_i},\alpha_e)\exp(-s_t(R_u)\sigma^2), \label{eq_ps_tn}
	\end{align}
in which the approximation sign comes from the using of upper bound of Gamma distribution,
\begin{align}
f_{1}(s,D_i,\alpha) &= \frac{1}{1+s\rho_t D_{i}^{-\alpha}},\nonumber\\
f_{2}(m,s,\eta,D_i,\alpha) &= \bigg(\frac{m}{m+s\eta\rho_u D_{i}^{-\alpha}}\bigg)^{m},\label{eq_f}
\end{align}
where $s_2(j,k,x) = k\beta_2(m_j)m_j g_j(x)$ and $s_t(x) = \frac{\theta}{ \rho_t}x^{\alpha_t}$, in which $g_l(x)=\frac{\theta}{\eta_l \rho_u}x^{\alpha_l}$, $g_n(x) = \frac{\theta}{\eta_n \rho_u}x^{\alpha_n}$ and $\beta_2(m) = (m!)^{-1/m}$ when $m>1$, otherwise $\beta_2(m) =  1$.
\label{lemma_successprob_1}
\end{lemma}
\begin{IEEEproof}
See Appendix \ref{app_proof_Ps}.
\end{IEEEproof}

The $b$-th moment is the final requirement of computing the SINR meta distribution, which is provided in the following theorem.
\begin{theorem}[$b$-th Moments] \label{theorem_Mb_1} The $b$-th moment of the conditional success probability is  given by
	\begin{align}
		M_b(\theta) &=  M_{b,l}(\theta)+M_{b,n}(\theta)+M_{b,tl}(\theta)+M_{b,tn}(\theta),\label{eq_mb}
	\end{align}
in which
	\begin{align}
		M_{b,l}(\theta) 
		&\approx\sum_{k_1=1}^{m_l}\sum_{k_2=1}^{m_l}\cdots\sum_{k_b=1}^{m_l}\binom{m_l}{k_1}\binom{m_l}{k_2}\cdots\binom{m_l}{k_b}(-1)^{\sum_{i=1}^{b}k_i+b}\int_{h}^{\sqrt{r_c^2+h^2}}P_l(\sqrt{x^2-h^2})\nonumber\\
		&\times\mathcal{A}_{\rm LoS,1}(x)\mathcal{L}_{ul}(s_2(l,k_1,x),s_2(l,k_2,x),\cdots,s_2(l,k_b,x),x)\frac{2x}{r_c^2}{\rm d}x,\label{eq_mb_l}\\
		M_{b,n}(\theta) &\approx \sum_{k_1=1}^{m_n}\sum_{k_2=1}^{m_n}\cdots\sum_{k_b=1}^{m_n}\binom{m_n}{k_1}\binom{m_n}{k_2}\cdots\binom{m_n}{k_b}(-1)^{\sum_{i=1}^{b}k_i+b}\int_{h}^{\sqrt{r_c^2+h^2}}P_n(\sqrt{x^2-h^2})\nonumber\\
		&\times\mathcal{A}_{\rm NLoS,1}(x)\mathcal{L}_{un}(s_2(n,k_1,x),s_2(n,k_2,x),\cdots,s_2(n,k_b,x),x)\frac{2x}{r_c^2}{\rm d}x,\label{eq_mb_n}\\
		M_{b,tl}(\theta) &\approx\int_{h}^{\sqrt{r_c^2+h^2}}  \int_{0}^{d_{lt}(y)}P_l(\sqrt{y^2-h^2})\tilde{\mathcal{L}}_{tl}(b,y)\frac{2y}{r_c^2}{\rm d}y,\label{eq_mb_tl}\\
		M_{b,tn}(\theta) 
		&\approx\int_{h}^{\sqrt{r_c^2+h^2}}  \int_{0}^{d_{nt}(y)}P_n(\sqrt{y^2-h^2})\tilde{\mathcal{L}}_{tn}(b,y)\frac{2y}{r_c^2}{\rm d}y\label{eq_mb_tn},
	\end{align}
	where the approximation sign comes from the using of upper bound of Gamma distribution, $\mathcal{A}_{\rm LoS,1}(x) = \bar{F}_{R_t}(d_{lt}(x))$ and $\mathcal{A}_{\rm NLoS,1}(x) = \bar{F}_{R_t}(d_{nt}(x))$, in which $\bar{F}_{ R_t}(x) = 1-F_{ R_t}(x)$ is the CCDF of $R_t$, and 
		\begin{align}
			\mathcal{L}_{{uj}}&(s_2(j,k_1,x),s_2(j,k_2,x),\cdots,s_2(j,k_b,x),x)\nonumber\\
			=& \exp\bigg(-2\pi\lambda_t\int_{d_{jt}(x)}^{\infty}\bigg[1-\prod_{i=1}^{b}f_1(s_2(j,k_i,x),z,\alpha_t)\bigg]z{\rm d}z\bigg)\nonumber\\
			& \times\prod_{e\in\{l,n\}}\exp\bigg(-2\pi\lambda_u\int_{h}^{\infty}\bigg[1-\prod_{i=1}^{b}f_{2}(m_e,s_2(j,k_i,x),\eta_e,z,\alpha_e)\bigg]z P_e(\sqrt{z^2-h^2}){\rm d}z\bigg)\nonumber\\
			&\times\exp\bigg(-\sum_{i=1}^{b} s_2(j,i,x)\sigma^2\bigg),\quad j\in\{l,n\},
		\end{align}
	\begin{align}
		&\tilde{\mathcal{L}}_{tj}(b,x)=\int_{0}^{d_{jt}(x)}f_{R_t}(r)f_{2}^{b}(m_j,s_t(r),\eta_{  j},r,\alpha_{ j})\mathcal{L}(b,s_t(r))\mathcal{L}_{t}(b,s_t(r),r){\rm d}r,\quad j\in\{l,n\},\\
		&\mathcal{L}_{t}(b,s,r)=\exp\bigg(-2\pi\lambda_t\int_{r}^{\infty}\bigg[1-f_{1}^{b}(s,z,\alpha_{t})\bigg]z{\rm d}z\bigg),\label{eq_laplace_lt}\\
		&\mathcal{L}(b,s)= \prod_{e\in\{l,n\}}\exp\bigg(-2\pi\lambda_{u,1}\int_{0}^{\infty}\bigg[1-f_{2}^{b}(m_e,s,\eta_e,z,\alpha_e)\bigg]\sqrt{z^2+h^2}P_e(z){\rm d}z\bigg)\exp(-bs\sigma^2).
	\end{align}
\end{theorem}
\begin{IEEEproof}
See Appendix \ref{app_proof_Mb}.
\end{IEEEproof}
\begin{remark}\label{rem_bth_moment}
As mentioned in  Remark \ref{rem_indicator}, we must use indicator functions to avoid the cross terms in (\ref{eq_mb}). If we use the association probabilities in (\ref{eq_mb}), in the case of the second moment, we will have $P_{s,l}P_{s,n}(\theta) = P_l(\sqrt{R_u^2-h^2})P_n(\sqrt{R_u^2-h^2})\mathcal{A}_{\rm LoS}(R_u)\mathcal{A}_{\rm NLoS}(R_u)g(I)$, where $g(I)$ is a function of interference terms. However, these cross terms are difficult to compute and meaningless. Once we use indicator functions: $\mathbbm{1}({\rm LoS})\mathbbm{1}({\rm LoS}) = \mathbbm{1}({\rm LoS})$ and $\mathbbm{1}({\rm LoS})\mathbbm{1}({\rm NLoS}) = 0$, which highly reduces the complexity of computing the high order moments.
\end{remark}

\begin{cor}[$b$-th moments of Noise-limited Scenario]\label{Cor_MCP}
If we only consider the noise and ignore the interference, (\ref{eq_mb_l}), (\ref{eq_mb_n}), (\ref{eq_mb_tl}) and (\ref{eq_mb_tn}) can be simplified,
\begin{align}
	M_{b,l}(\theta) 
	&=\int_{h}^{\sqrt{r_c^2+h^2}}P_l(\sqrt{x^2-h^2})\mathcal{A}_{\rm LoS,1}(x)\exp(-b g_l(x)\sigma^2)\frac{2x}{r_c^2}{\rm d}x,\label{eq_mb_l1}\\
	M_{b,n}(\theta) &=\int_{h}^{\sqrt{r_c^2+h^2}}P_n(\sqrt{x^2-h^2})\mathcal{A}_{\rm NLoS,1}(x)\exp(-b g_n(x)\sigma^2)\frac{2x}{r_c^2}{\rm d}x,\label{eq_mb_n1}\\
	M_{b,tl}(\theta) &=\int_{h}^{\sqrt{r_c^2+h^2}}  \int_{0}^{d_{lt}(r)}P_l(\sqrt{y^2-h^2})\exp(-bs_t(r)\sigma^2)f_{R_t}(r)\frac{2y}{r_c^2}{\rm d}r{\rm d}y,\label{eq_mb_tl1}\\
	M_{b,tn}(\theta) 
&=\int_{h}^{\sqrt{r_c^2+h^2}}  \int_{0}^{d_{nt}(r)}P_n(\sqrt{y^2-h^2})\exp(-bs_t(r)\sigma^2)f_{R_t}(r)\frac{2y}{r_c^2}{\rm d}r{\rm d}y\label{eq_mb_tn1}.
\end{align}
\end{cor}
\begin{IEEEproof}
	Proof completes by ignoring the Laplace transform of the interference term in (\ref{eq_mb_l}), (\ref{eq_mb_n}), (\ref{eq_mb_tl}) and (\ref{eq_mb_tn}).
	\end{IEEEproof}

\subsection{PPP Distributed Users}
In this part, we provide the analysis for the system model 2: the locations of users are modeled by PPP. Compared with the first system model, the first contact distances of UAVs are different. Hence, our analysis start from the first contact distances of UAVs, which are shown in the below lemma.
\begin{lemma}[Distance Distribution] The probability density function of the distances between the typical user and  the nearest available NLoS and LoS UAV, denoted by $f_{ R_{ n}}(r)$ and $f_{ R_{ l}}(r)$, respectively, are given in \cite{alzenad2019coverage} as
	\begin{align}
		f_{ R_{ n}}(r) &= 2\pi\lambda_{ u,2}P_{ n}(\sqrt{r^2-h^2})r\exp\bigg(-2\pi\lambda_{ u,2}\int_{0}^{\sqrt{r^2-h^2}}zP_{\rm n}(z){\rm d}z\bigg),\nonumber\\
		f_{ R_{ l}}(r) &= 2\pi\lambda_{ u,2}P_{ l}(\sqrt{r^2-h^2})r\exp\bigg(-2\pi\lambda_{ u,2}\int_{0}^{\sqrt{r^2-h^2}}zP_{ l}(z){\rm d}z\bigg)\label{dist_u_p_l},
	\end{align}
	where $P_{ n}(r)$ and $P_{ l}(r)$ are defined in (\ref{pl_pn}). Therefore, the CDFs of $R_n$ and $R_l$ are
	\begin{align}
	F_{ R_{ n}}(r) &= 1-\exp\bigg(-2\pi\lambda_{ u,2}\int_{0}^{\sqrt{r^2-h^2}}zP_{ n}(z){\rm d}z\bigg),\nonumber\\
	F_{ R_{ l}}(r) &= 1-\exp\bigg(-2\pi\lambda_{ u,2}\int_{0}^{\sqrt{r^2-h^2}}zP_{ l}(z){\rm d}z\bigg)\label{dist_u_p_l}.
	\end{align}
\end{lemma}
\begin{lemma}[Nearest Interfering BSs] When the reference user is associated with the NLoS/LoS UAVs, the nearest interfering LoS/NLoS UAVs are at least $d_{nl}(R_n)$ and $d_{nl}(R_l)$ away, respectively. When the reference user is associated with the TBS, the nearest interfering LoS/NLoS UAVs are at least $d_{tl}(R_n)$ and $d_{tl}(R_l)$ away, respectively. All of the nearest interfering BSs are given as, respectively,
\begin{align}
	d_{ln}(R_l) &= \max(h,(\frac{\eta_n}{\eta_l})^{\frac{1}{\alpha_n}}R_{l}^{\frac{\alpha_l}{\alpha_n}}),\nonumber\\
	d_{nl}(R_n) &= \max(h,(\frac{\eta_l}{\eta_n})^{\frac{1}{\alpha_l}}R_{n}^{\frac{\alpha_n}{\alpha_l}}),\nonumber\\
	d_{tl}(R_t) &= \max(h,(\frac{\rho_u\eta_l}{\rho_t})^{\frac{1}{\alpha_l}}R_{t}^{\frac{\alpha_t}{\alpha_l}}),\nonumber\\
	d_{tn}(R_t) &= \max(h,(\frac{\rho_u\eta_n}{\rho_t})^{\frac{1}{\alpha_n}}R_{t}^{\frac{\alpha_t}{\alpha_n}}).
\end{align}
Here, the distances of the nearest interfering TBS in the case that associated  LoS/NLoS UAVs are the same as system model 1, thus omitted here.
\end{lemma}

Recall that the user associates with the nearest LoS/NLoS UAV or the nearest TBS which provides the strongest average received power, the SINR is given as
\begin{align}
	{\rm SINR} = \frac{\max{(p_u,p_t)}}{I+\sigma^2},
\end{align}
 the aggregate interference in system model 2 is
\begin{align}
	I  =& \sum_{N_i\in\Phi_{ u_n,2}}\eta_{ n}\rho_{ u}G_{ n,N_i}D_{ N_i}^{-\alpha_{ n}}+\sum_{L_j\in\Phi_{ u_l,2}}\eta_{ l}\rho_{ u}G_{ l,L_j}D_{ L_j}^{-\alpha_{ l}}+\sum_{T_k\in\Phi_{ t}\setminus \{b_0\}}\rho_{ t}H_{T_k}D_{ T_k}^{-\alpha_{ t}},\nonumber
\end{align}
where $\Phi_{ u_n,2}$ and $\Phi_{ u_l,2}$ are subsets of $\Phi_{ u,2}\setminus \{b_0\}$ denote the locations of NLoS and LoS UAVs, respectively, and $b_0$ is the location of the serving BS, which can be either UAV or TBS.
Consequently, the conditional success probability of the reference link is given by
\begin{align}
	&P_s(\theta)= \mathbb{P}({\rm SINR}>\theta|\Phi_{  t},\Phi_{u,2},\Phi_{\rm user})= \mathbb{P}\bigg(\frac{\max{(p_u,p_t)}}{I+\sigma^2}>\theta|\Phi_{  t},\Phi_{u,2},\Phi_{\rm user}\bigg)  \nonumber\\
	&= \mathbbm{1}({\rm LoS})\mathbb{P}\bigg(\frac{p_{u,l}}{I_{l}+\sigma^2}>\theta\bigg) +\mathbbm{1}({\rm NLoS})\mathbb{P}\bigg(\frac{p_{u,n}}{I_{n}+\sigma^2}>\theta\bigg)+\mathbbm{1}({\rm TBS})\mathbb{P}\bigg(\frac{p_{t}}{I_{t}+\sigma^2}>\theta\bigg)\nonumber\\
	&= \mathbbm{1}(p_l > p_n)\mathbbm{1}(p_l > p_t)\mathbb{P}\bigg(\frac{p_{u,l}}{I_{l}+\sigma^2}>\theta\bigg) +\mathbbm{1}(p_n > p_l)\mathbbm{1}(p_n > p_t)\mathbb{P}\bigg(\frac{p_{u,n}}{I_{n}+\sigma^2}>\theta\bigg)\nonumber\\
	&\quad+\mathbbm{1}(p_t > p_n)\mathbbm{1}(p_t > p_l)\mathbb{P}\bigg(\frac{p_{t}}{I_{t}+\sigma^2}>\theta\bigg)\nonumber\\
	&= P_{s,l}(\theta) +P_{s,n}(\theta)+P_{s,t}(\theta),
\end{align}
where $\mathbbm{1}(p_l > p_n)\mathbbm{1}(p_l > p_t)$ denotes the average received power from the nearest LoS UAV is stronger than that of the nearest NLoS UAV and TBS, hence, it denotes that the user associates with the LoS UAV. Similar meaning applies to $\mathbbm{1}(p_n > p_l)\mathbbm{1}(p_n > p_t)$ and $\mathbbm{1}(p_t > p_n)\mathbbm{1}(p_t > p_l)$.

\begin{lemma}[Conditional Success Probability] The conditional success probability of the reference link is given by
	\begin{align}
		 P_{s,l}(\theta)&\approx 	\sum_{k_l=1}^{m_l}\binom{m_l}{k_l}(-1)^{k_l+1}\mathbbm{1}(p_l > p_n)\mathbbm{1}(p_l > p_t)\prod_{T_k\in\Phi_{ t}}f_1(s_2(l,k_l,R_l),D_{T_k},\alpha_t)\nonumber\\
		&\prod_{e\in\{l,n\}}\prod_{u_i\in\Phi_{ u_e,2}}f_2(m_e,s_2(l,k_l,R_l),\eta_l,D_{u_e},\alpha_e)\exp(-s_2(l,k_l,R_l)\sigma^2),\\
		 P_{s,n}(\theta)&\approx	\sum_{k_n=1}^{m_n}\binom{m_n}{k_n}(-1)^{k_n+1}\mathbbm{1}(p_n > p_l)\mathbbm{1}(p_n > p_t)\prod_{T_k\in\Phi_{ t}}f_1(s_2(n,k_n,R_n),D_{T_k},\alpha_t)\nonumber\\
		 &\prod_{e\in\{l,n\}}\prod_{u_i\in\Phi_{ u_e,2}}f_2(m_e,s_2(n,k_n,R_n),\eta_n,D_{u_e},\alpha_e)\exp(-s_2(n,k_n,R_n)\sigma^2),\\
		P_{s,t}(\theta) &\approx \mathbbm{1}(p_t > p_n)\mathbbm{1}(p_t > p_l)\prod_{T_k\in\Phi_{ t}\setminus \{b_0\}}f_1(s_t(R_t),D_{T_k},\alpha_t)\nonumber\\
	 &\prod_{e\in\{l,n\}}\prod_{u_i\in\Phi_{ u_e,2}}f_2(m_e,s_t(R_t),\eta_l,D_{u_e},\alpha_e)\exp(-s_t(R_t)\sigma^2),
	\end{align}
	where $u_0$ and $t_0$ are the serving UAV or TBS.
\end{lemma}
\begin{theorem}[$b$-th Moments] The $b$-th moment of the conditional success probability of TBS-only network is given by
\begin{align}
	M_b(\theta) 
	&= M_{b,l}(\theta)+M_{b,n}(\theta)+M_{b,t}(\theta),
\end{align}
where
	\begin{align}
		M_{b,l}(\theta) 
		&\approx\sum_{k_1=1}^{m_l}\sum_{k_2=1}^{m_l}\cdots\sum_{k_b=1}^{m_l}\binom{m_l}{k_1}\binom{m_l}{k_2}\cdots\binom{m_l}{k_b}(-1)^{k_1+k_2+\cdots+k_b+b},\label{eq_Mbl_PPP}\nonumber\\
		&\int_{h}^{\infty}\mathcal{A}_{\rm LoS,2}(x)f_{ R_l}(x)\mathcal{L}_{ul,2}(s_2(l,k_1,x), s_2(l,k_2,x), \cdots, s_2(l,k_b,x),x){\rm d}x,\\
		M_{b,n}(\theta)  &\approx\sum_{k_1=1}^{m_n}\sum_{k_2=1}^{m_n}\cdots\sum_{k_b=1}^{m_n}\binom{m_n}{k_1}\binom{m_n}{k_2}\cdots\binom{m_n}{k_b}(-1)^{k_1+k_2+\cdots+k_b+b}\nonumber\\
		&\int_{h}^{\infty}\mathcal{A}_{\rm NLoS,2}(x)f_{ R_n}(x)\mathcal{L}_{un,2}(s_2(n,k_1,x), s_2(n,k_2,x), \cdots, s_2(n,k_b,x),x){\rm d}x,\label{eq_Mbn_PPP}\\
		M_{b,t}(\theta) &=\int_{0}^{\infty}\mathcal{A}_{\rm TBS}(x)f_{ R_t}(x)\mathcal{L}_2(b,s_t(x),x){\rm d}x,\label{eq_Mbt_PPP}
	\end{align}
 where the approximation sign comes from the using of upper bound of Gamma distribution, $\mathcal{A}_{\rm LoS,2}(x)$, $\mathcal{A}_{\rm NLoS,2}(x)$ and $\mathcal{A}_{\rm TBS,2}(x)$ are commonly known as association probabilities, which are derived by taking the expectation of the indicator functions: $\mathcal{A}_{\rm LoS,2}(x) = \bar{F}_{ R_t}(d_{lt}(x))\bar{F}_{ R_{R_n}}(d_{ln}(x))$, $\mathcal{A}_{\rm NLoS,2}(x) = \bar{F}_{ R_t}(d_{nt}(x))\bar{F}_{ R_{R_l}}(d_{nl}(x))$ and $\mathcal{A}_{\rm TBS,2}(x) = \bar{F}_{ R_l}(d_{tl}(x))\bar{F}_{ R_{R_n}}(d_{tn}(x))$,
	\begin{align}
		&\mathcal{L}_{\{ul,un\},2}(s_2(\{l,n\},k_1,x), s_2(\{l,n\},k_2,x), \cdots, s_2(\{l,n\},k_b,x),x)\nonumber\\
		&=\exp\bigg(-2\pi\lambda_t\int_{\{d_{lt}(x),d_{nt}(x)\}}^{\infty}\bigg[1-\prod_{i=1}^{b}f_1(s_2(\{l,n\},k_i,x),z,\alpha_t)\bigg]z{\rm d}z\bigg)\nonumber\\
		& \exp\bigg(-2\pi\lambda_u\int_{\{d_{ln}(x),x\}}^{\infty}\bigg[1-\prod_{i=1}^{b}f_{2}(m_n,s_2(\{l,n\},k_i,x),\eta_n,z,\alpha_n)\bigg]zP_n(\sqrt{x^2-h^2}){ d}z\bigg)\nonumber\\
		&\exp\bigg(-2\pi\lambda_u\int_{\{x,d_{nl}(x)\}}^{\infty}\bigg[1-\prod_{i=1}^{b}f_{2}(m_l,s_2(\{l,n\},k_i,x),\eta_l,z,\alpha_l)\bigg]zP_l(\sqrt{x^2-h^2}){\rm d}z\bigg)\nonumber\\
		&\exp(-\sum_{i = 1}^{b}s_2(\{l,n\},k_i,x)\sigma^2),
	\end{align}
and
\begin{align}
	\mathcal{L}_2(b,s,x) &=\exp\bigg(-2\pi\lambda_{u,2}\int_{a(x)}^{\infty}\bigg[1-f_{2}^{b}(m_n,s,\eta_n,z,\alpha_{ n})\bigg]\sqrt{z^2+h^2}P_n(z){\rm d}z\bigg)\nonumber\\
	& \exp\bigg(-2\pi\lambda_{u,2}\int_{b(x)}^{\infty}\bigg[1-f_{2}^{b}(m_l,s,\eta_l,z,\alpha_{ l})\bigg]\sqrt{z^2+h^2}P_l(z){\rm d}z\bigg)\nonumber\\
	&\exp\bigg(-2\pi\lambda_t\int_{c(x)}^{\infty}\bigg[1-f_{1}^{b}(s,z,\alpha_{ t})\bigg]z{\rm d}z\bigg)\exp(-b s(x)\sigma^2),\nonumber
\end{align}
in which
		\begin{align}
	a(x)=\left\{ 
	\begin{aligned}
		x,  & \quad \text{\rm in the case of} \quad M_{b,n},\\
		d_{tn}(x),  & \quad \text{\rm  in the case of}\quad  M_{b,t},\\
	\end{aligned} \right.\nonumber
\end{align}
\begin{align}
	b(x)=\left\{ 
	\begin{aligned}
		d_{nl}(x),  & \quad \text{\rm in the case of} \quad M_{b,n},\\
		d_{tl}(x),  & \quad \text{\rm in the case of} \quad M_{b,t},\\
	\end{aligned} \right.\nonumber
\end{align}
\begin{align}
	c(x)=\left\{ 
	\begin{aligned}
		d_{nt(x)},  & \quad \text{\rm in the case of} \quad M_{b,n},\\
		x,  & \quad \text{\rm in the case of} \quad M_{b,t}.\\
	\end{aligned} \right.\nonumber
\end{align}
\end{theorem}
\begin{IEEEproof}
	Similar to the proof of Theorem \ref{theorem_Mb_1}, thus omitted here.
	\end{IEEEproof}
\begin{cor}[$b$-th moments of Noise-limited Scenario]\label{Cor_PPP}
	If we only consider the noise and ignore the interference, (\ref{eq_Mbl_PPP}), (\ref{eq_Mbn_PPP})  and (\ref{eq_Mbt_PPP}) can be simplified,
	\begin{align}
		M_{b,l}(\theta) 
		&=\int_{h}^{\infty}\mathcal{A}_{\rm LoS,2}(x)\exp(-b g_l(x)\sigma^2)f_{ R_l}(x){\rm d}x,\label{eq_Mbl_PPP1}\\
		M_{b,n}(\theta) &=\int_{h}^{\infty}\mathcal{A}_{\rm NLoS,2}(x)\exp(-b g_n(x)\sigma^2)f_{ R_n}(x){\rm d}x,\label{eq_Mbn_PPP1}\\
		M_{b,tl}(\theta) &=\int_{0}^{\infty}  \mathcal{A}_{\rm TBS,2}(x)\exp(-bs_t(r)\sigma^2)f_{R_t}(r){\rm d}r,\label{eq_Mbt_PPP1}
	\end{align}
\end{cor}
\begin{IEEEproof}
	Proof completes by ignoring the Laplace transform of the interference term in (\ref{eq_Mbl_PPP}), (\ref{eq_Mbn_PPP}), and (\ref{eq_Mbt_PPP}).
\end{IEEEproof}

\subsection{TBS Only Networks}
We are interested in the improvement of the reliability of deploying UAVs in the wireless networks. Therefore, we need to compute the meta distribution of TBS-only networks. Based on the nearest BS association policy, the SINR at the user is given by
\begin{align}
	{\rm SINR} = \frac{p_t}{I+\sigma^2},
\end{align}
where $\sigma^2$ is thermal noise with Gaussian distribution, and the aggregate interference is 
\begin{align}
	I  =&  \sum_{T_k\in\Phi_{ t}\setminus \{t_0\}}\rho_{ t}H_{T_k}D_{ T_k}^{-\alpha_{ t}},\nonumber
\end{align}
where $t_0$ is the location of the serving TBS.
\begin{lemma}[Conditional Success Probability]The conditional success probability of the reference link in TBS-only networks is given by
\begin{align}
	\mathbb{P}_s(\theta) &= \mathbb{P}({\rm SINR}>\theta|\Phi_{  t},\Phi_{\rm user})= \mathbb{P}\bigg(\frac{p_t}{I+\sigma^2}>\theta|\Phi_{  t},\Phi_{\rm user}\bigg)=\exp(-s_t(R_t)(I_{t}+\sigma^2))\nonumber\\
	&= \prod_{T_k\in\Phi_{ t}\setminus \{t_0\}}f_{1}^{b}(s_t(R_t),D_{T_k},\alpha_{ t}) \exp(-s_t(R_t)\sigma^2),
\end{align}
where $s_t(x) = \frac{\theta}{ \rho_t}x^{\alpha_t}$.
\end{lemma}

Consequently, we obtain the moments of the TBS-only networks.

\begin{theorem}[$b$-th Moments] The $b$-th moment of the conditional success probability is given by
	\begin{align}
		M_b(\theta) &= \int_{0}^{\infty}\mathcal{L}_t(b,s_t(x),x)\exp(-bs_t(x)\sigma^2)f_{R_{t}}(x){\rm d}x,\label{eq_Mb_TBSonly}
	\end{align}
where $\mathcal{L}_t(b,s_t(x),x)$ is defined in (\ref{eq_laplace_lt}).
\end{theorem}
\begin{IEEEproof}
	 The $b$-th moment of the conditional success probability is derived by taking the expectation over the locations, given by
	\begin{align}
		M_b(\theta) &= \mathbb{E}[\mathbb{P}^b_s(\theta)] = \mathbb{E}\bigg[\prod_{T_k\in\Phi_{  t}}f_{1}^{b}(s_t(R_t),D_{T_k},\alpha_{ t})\exp(-b s_t(R_t)\sigma^2)\bigg]\nonumber\\
		&= \mathbb{E}\bigg[\exp\bigg(-2\pi\lambda_t\int_{R_t}^{\infty}\bigg[1-f_{1}^{b}(s_t(R_t),z,\alpha_{ t})\bigg]z{\rm d}z\bigg)\exp(-bs_t(R_t)\sigma^2)\bigg],
	\end{align}
proof completes by taking the expectation over the distance to the serving BS.
\end{IEEEproof}
\begin{cor}[$b$-th moments of Noise-limited Scenario]\label{Cor_TBS}
	If we only consider the noise and ignore the interference, (\ref{eq_Mb_TBSonly}) can be simplified,
\begin{align}
	M_b(\theta) &= \int_{0}^{\infty}\exp(-bs_t(x)\sigma^2)f_{R_{t}}(x){\rm d}x,\label{eq_Mb_TBSonly1}
\end{align}
\end{cor}
\begin{IEEEproof}
	Proof completes by ignoring the Laplace transform of the interference term in (\ref{eq_Mb_TBSonly}).
\end{IEEEproof}

To fairly compare TBS-only and UAV-assisted scenarios, we also analyze the case by converting UAVs into ground BSs to keep the same intensity of transmitters, say scenario 1. Note that we have both user location models for scenario 1, MCP and PPP, and in each figure in Section-IV, we use the same user distribution model. The results of scenario 1 are obtained by setting the altitude of UAVs $h = 0$ m and $\eta_{ l}=\eta_{ n} = 0$ dB. However, we would like to mention that this scenario is not practical owing to the dynamic of the locations of user clusters and the difficulties of high dense deployment of TBSs, e.g., the UAVs can adjust and optimize their locations, for instance, deploying in the cluster centers, while the fixed TBSs cannot.


\subsection{Meta Distribution of UAV-assisted Networks}
Recall that the meta distribution is the CCDF of the conditional success probability, which describes the reliability of the network, e.g., the probability that the reference link exceeds that reliability threshold $\theta$. We first give the exact expression of the meta distribution which is generally solved by using the Gil-Pelaez theorem \cite{haenggi2015meta}, \cite{gil1951note}. 

The exact expression is given by
	\begin{align}
		\bar{F}_{P_s}(\theta,\gamma) &=  \frac{1}{2}+\frac{1}{\pi}\int_{0}^{\infty}\frac{\Im(\exp(-jt\log \gamma)M_{jt}(\theta))}{t}{\rm d}t,\label{eq_MetaExact}
	\end{align}
	where $M_{jt}(\theta)$ is the ${jt}$-th moment of the conditional success probability $P_s(\theta)$, $j$ is the imaginary unit. $M_{jt}(\theta)$ is computed by replacing the $b$ in $M_b(\theta)$ that we derived in previous sections, and $\Im(\cdot)$ is the imaginary part of a complex number.

However, computing the meta distribution using (\ref{eq_MetaExact}) is difficult since it requires computing the imaginary moments. Alternatively, we use the beta approximation, which is given in the following remark. The beta approximation shows great matching for large range of $\theta$ in TBSs related networks, such as \cite{haenggi2015meta,haenggi2021meta1,haenggi2021meta2}, and it only requires to compute the first and the second moments.
\begin{remark}[Beta Approximation]
	Using the beta approximation, the meta distribution can be approximated as
	\begin{align}
		\bar{F}^{'}_{P_s}(\theta,\gamma)	\approx 1-I_{\gamma}\bigg(\frac{M_1(\theta)(M_1(\theta)-M_2(\theta))}{M_2(\theta)-M_1^2(\theta)},\frac{(M_1(\theta)-M_2(\theta))(1-M_1(\theta))}{M_2(\theta)-M_1^2(\theta)}\bigg),\label{eq_MetaBetaApp}
	\end{align}
	where,
	\begin{align}
		I_x(a,b) = \frac{\int_{0}^{x}t^{a-1}(1-t)^{b-1}{\rm d}t}{B(a,b)},
	\end{align}
	and $B(a,b) = \int_{0}^{1}t^{a-1}(1-t)^{b-1}{\rm d}t$.
\end{remark}
As shown, beta approximation only requires to compute the first and the second moment, which highly reduces the computing complexity.

\section{Numerical Results}
	\begin{table}[ht]\caption{Table of Parameters}\label{par_val}
	\centering
	\begin{center}
		\resizebox{0.7\columnwidth}{!}{
			\renewcommand{\arraystretch}{1}
			\begin{tabular}{ {c} | {c} | {c}  }
				\hline
				\hline
				\textbf{Parameter} & \textbf{Symbol} & \textbf{Simulation Value}  \\ \hline
				Density of TBSs & $\lambda_{ t}$ & $1$ km$^{-2}$ \\ \hline
				Density of UAVs & $\lambda_{ u,\{1,2\}}$ & $1$ km$^{-2}$ \\\hline
				UAV altitude & $h$ & 100 m\\\hline
				Radius of MCP disk & $r_c$ & 100 m \\\hline
				Environment parameters (highrise urban) & $P_{l,1} (e_1,e_2)$ & $(27,0.08)$ \cite{al2014optimal} \\\hline
				Environment parameters (dense urban) & $P_{l,2} (e_1,e_2)$ & $(12,0.11)$ \cite{al2014optimal} \\\hline
				Environment parameters (urban)& $P_{l,3} (e_1,e_2)$ & $(9.6,0.16)$ \cite{al2014optimal}  \\\hline
				Environment parameters (suburban)& $P_{l,4} (e_1,e_2)$ & $(4.88,0.43)$ \cite{al2014optimal}  \\\hline
				Transmission power of UAVs and TBSs& $\rho_{ u}$, $\rho_{ t}$ & 0.2 W, 10 W\\\hline
				Noise power & $\sigma^2 $ & $10^{-9}$ W\\\hline
				N/LoS UAV, TBS path-loss exponent & $\alpha_{ n},\alpha_{ l},\alpha_{ t}$ & $4,2.1,4$ \\\hline
				N/LoS fading parameters & $m_{ n},m_{ l}$ & $1,3$ \\\hline
				N/LoS additional loss& $\eta_{ n},\eta_{ l}$ & $-20,0$ dB 
				\\\hline\hline
		\end{tabular}}
	\end{center}
\end{table}
In this section, we validate the theoretical results via Monte Carlo simulations with a large number of iterations to ensure accuracy. We would like to mention that, even though we have two approximations in this work, one is using the upper bound of the Gamma function and one is using beta approximation, all the simulation results match well with the analysis results.

For the first system model, we first generate two independent PPPs for the locations of user cluster centers, and TBSs, respectively, and UAVs are located above the cluster centers. We then generate the locations of users, which are uniformly distributed within the user clusters, and the number of users is a Poisson random variable. Considering the users near the edge of the area of interest, the simulation windows of BSs are much larger than the simulation window of user clusters. While the realizations of all the locations are fixed, the fading realizations change in each iteration, and we compute the conditional success probability for each user link. Finally, we obtain the CCDF of the conditional success probability of a fixed realization. The simulation of the second system model follows a similar way by modeling the locations of users, UAVs, and TBSs by three independent PPPs directly under the same simulation windows. In the simulation and analysis of scenario 1, the altitude of UAVs is set at $0$ and $\eta_{ l}=\eta_{ n} = 0$ dB, and in this way, all the aerial BSs are converted into ground BSs. Unless stated otherwise, we use the system parameters listed herein \ref{par_val}. Besides, we would like to clarify that even though the equations look long and complex, they are not difficult in computing and the time for each point is about 30 min. The complexity and time consumption of the numerical computation of the proposed expressions may be significantly reduced in the special case of $m_n=m_l=1$ by following the approximation proposed in \cite{Approx} (however, the spatial distribution of interferers should be carefully computed). Finally, we assume that TBSs and UAVs transmit their signals at $2$ GHz with a system bandwidth of $10$ MHz.

\begin{figure}
	\centering
	\subfigure[]{\includegraphics[width=0.32\columnwidth]{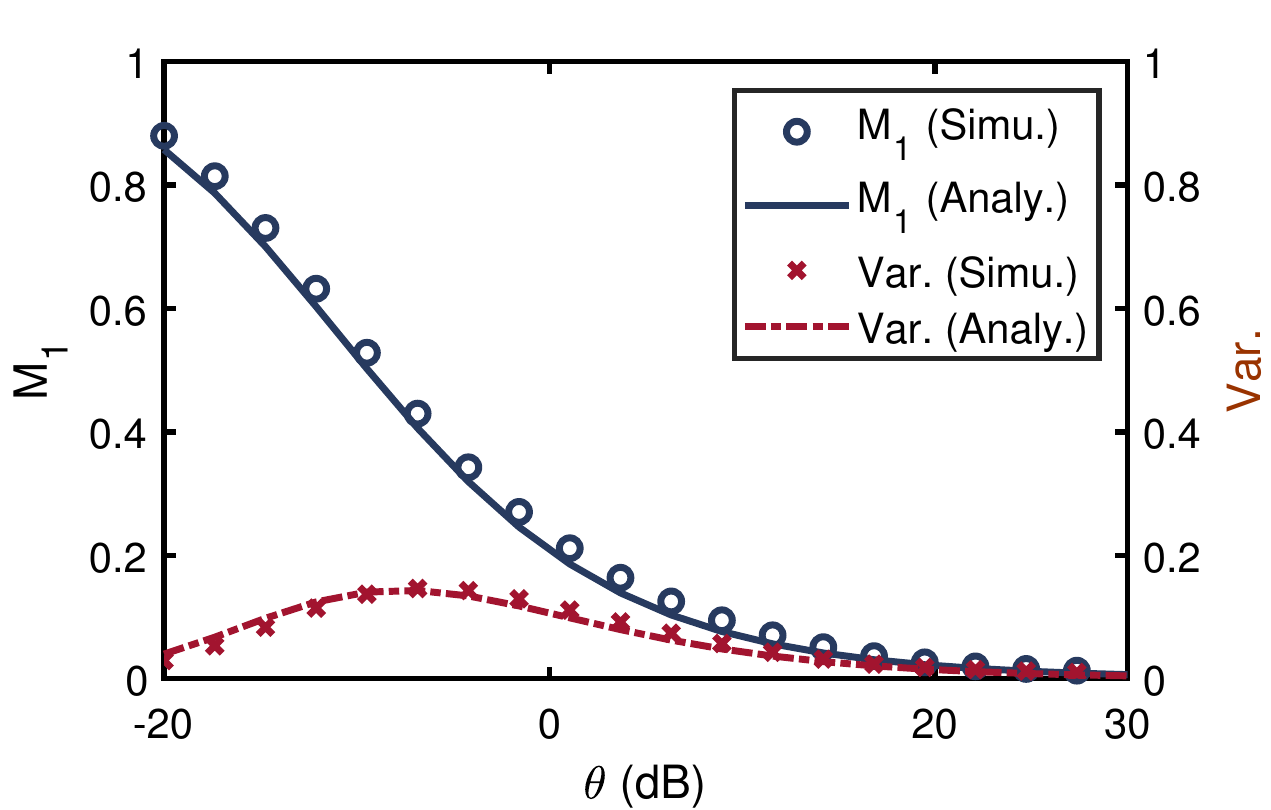}}
	\subfigure[]{\includegraphics[width=0.32\columnwidth]{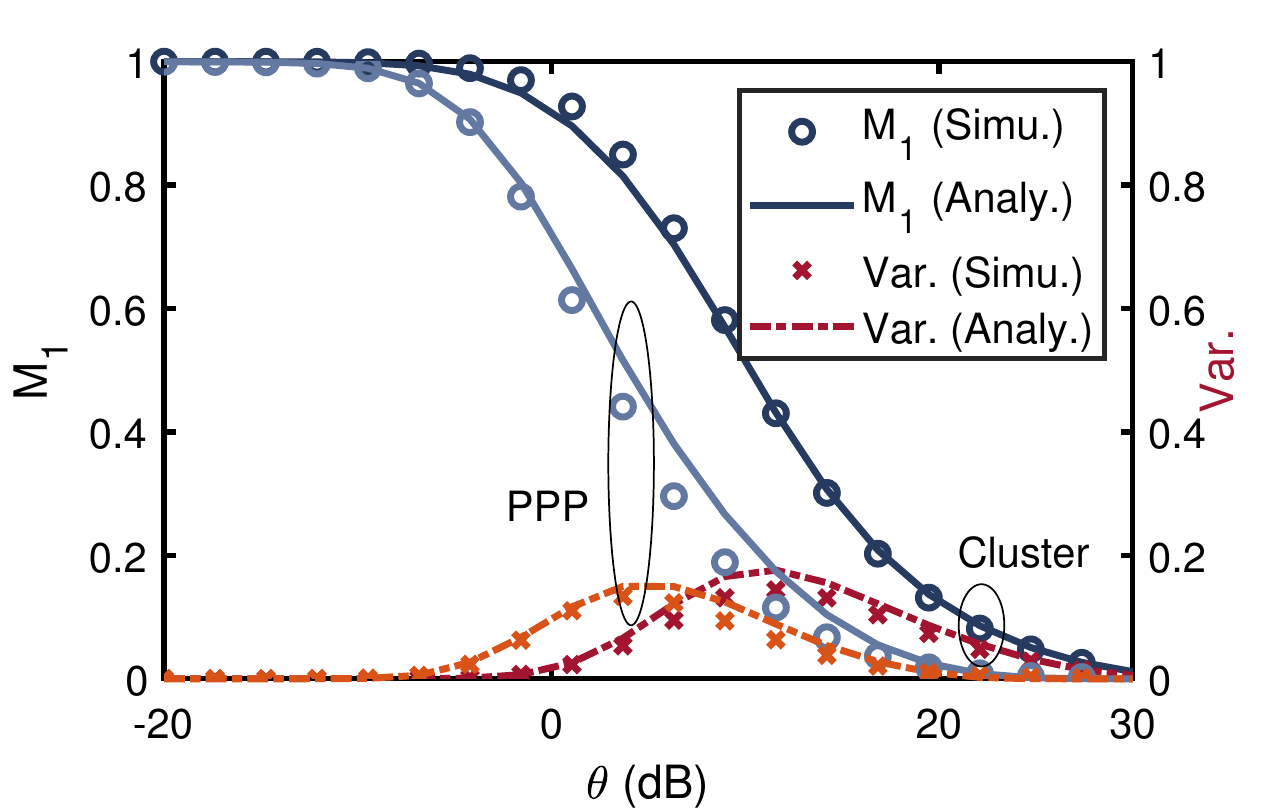}}
	\subfigure[]{\includegraphics[width=0.32\columnwidth]{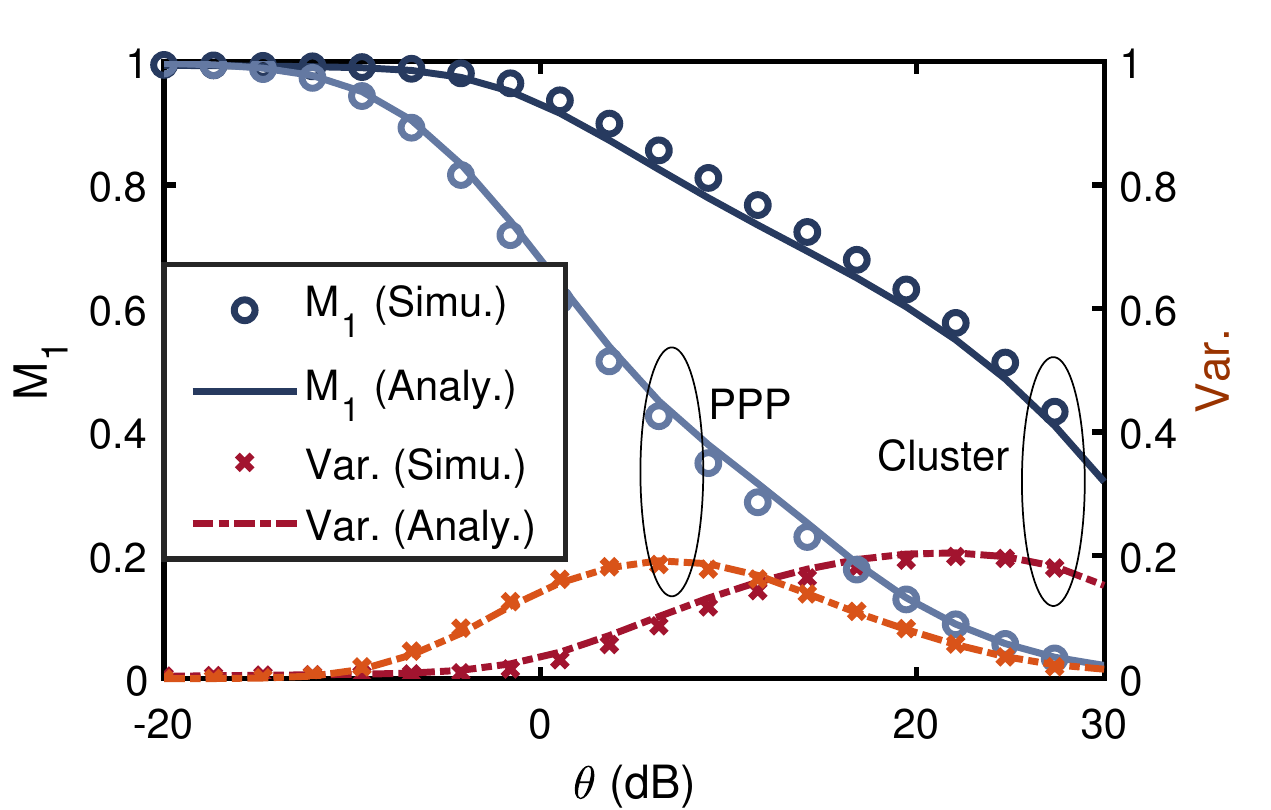}}
	\subfigure[]{\includegraphics[width=0.33\columnwidth]{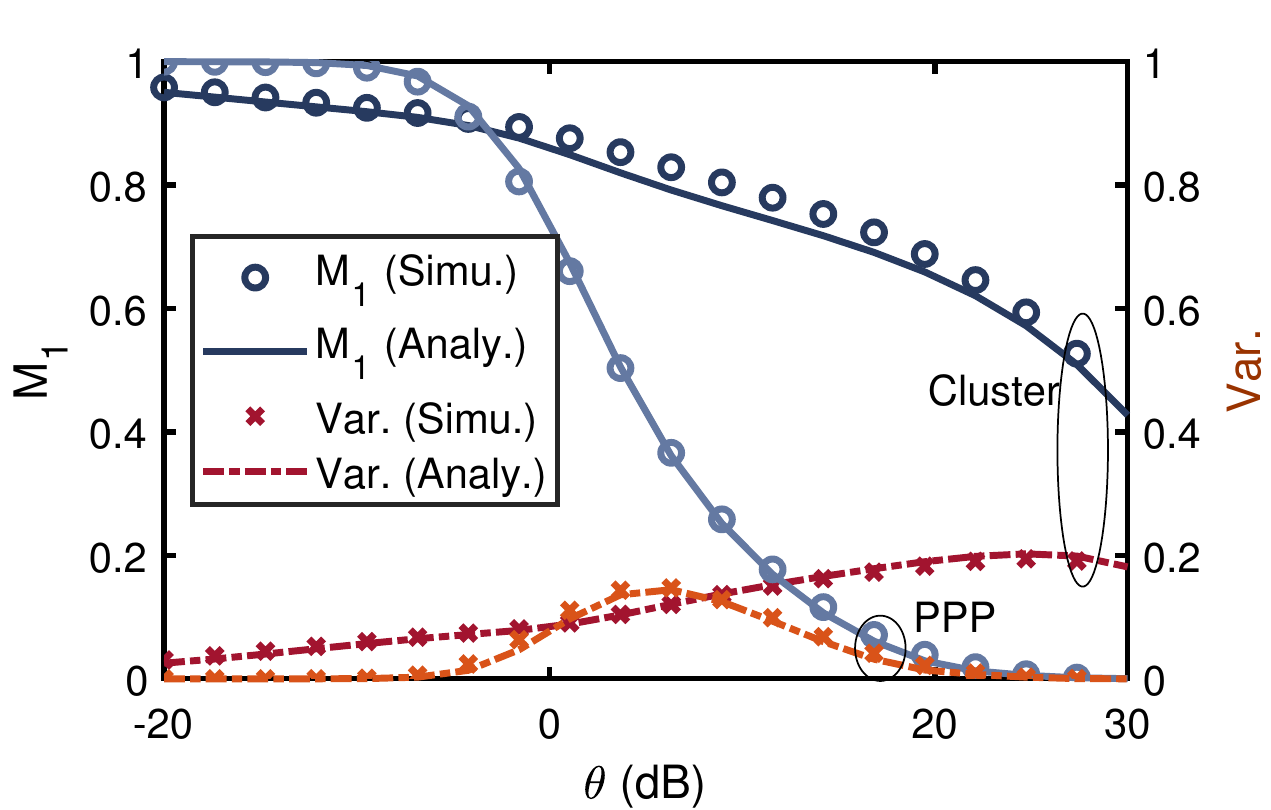}}
	\subfigure[]{\includegraphics[width=0.33\columnwidth]{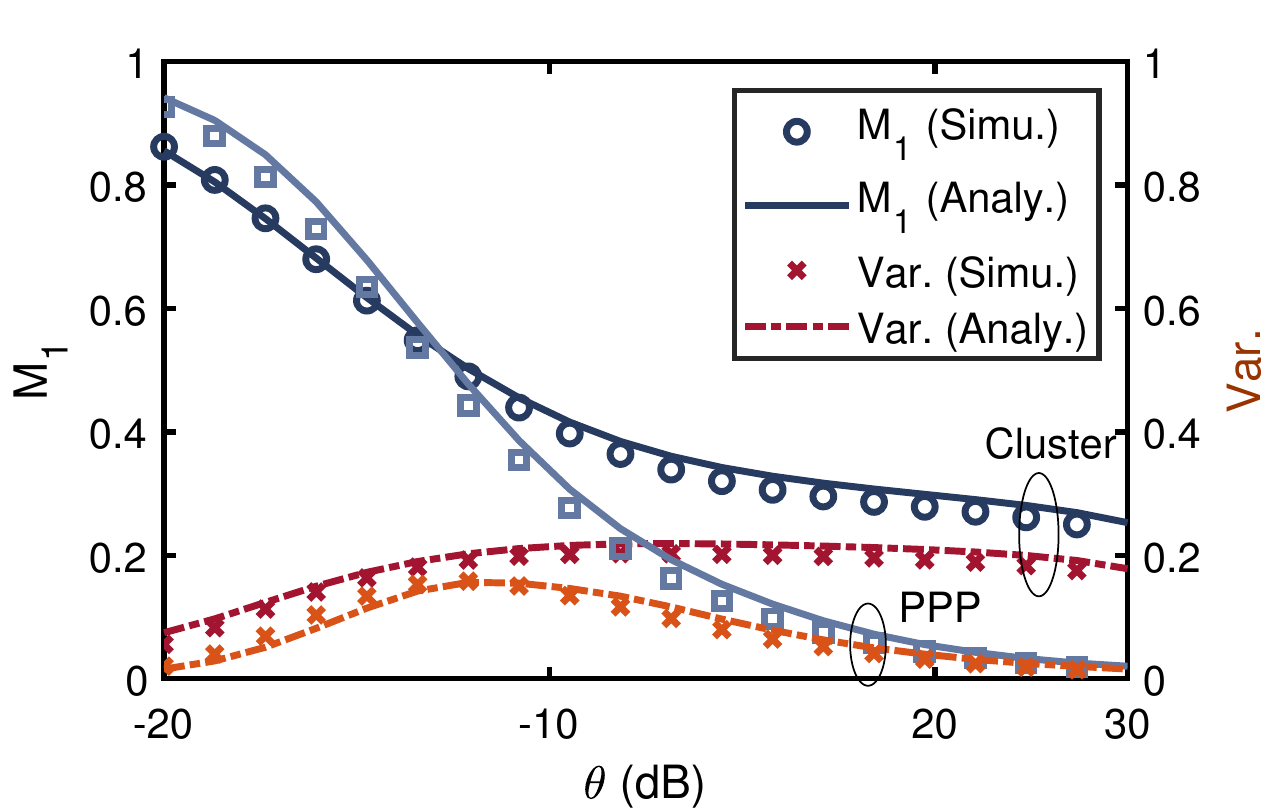}}
	\caption{The simulation and analysis results of the first moment  and the variance of the two system models. The first moment $M_1$ of the conditional success probability and the variance $M_2-M_1^2$ in \textbf{(a)} TBS-only cellular network, \textbf{(b)} UAV-assisted cellular networks in suburban areas ($e_1 = 4.88, e_2 = 0.43$), \textbf{(c)} UAV-assisted cellular networks in urban areas ($e_1 = 9.6, e_2 = 0.16$), \textbf{(d)} UAV-assisted cellular networks in dense urban areas ($e_1 = 12, e_2 = 0.11$), \textbf{(e)} UAV-assisted cellular networks in highrise urban areas ($e_1 = 27, e_2 = 0.08$).}
	\label{Fig_meanvar}
\end{figure}
\begin{figure}
	\subfigure[$\theta = -10$ dB.]{\includegraphics[width=0.5\columnwidth]{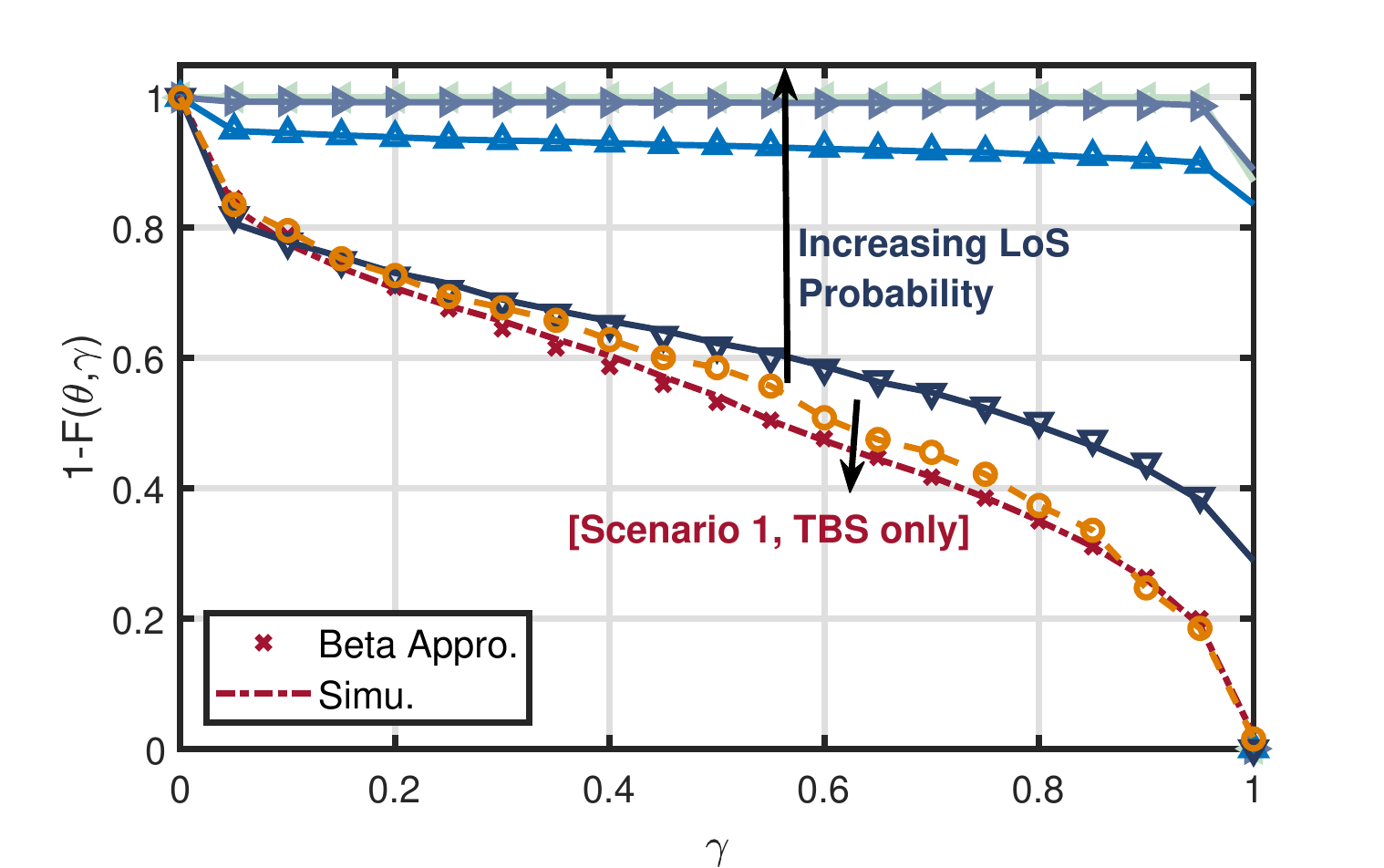}}
	\subfigure[$\theta = 0$ dB.]{\includegraphics[width=0.5\columnwidth]{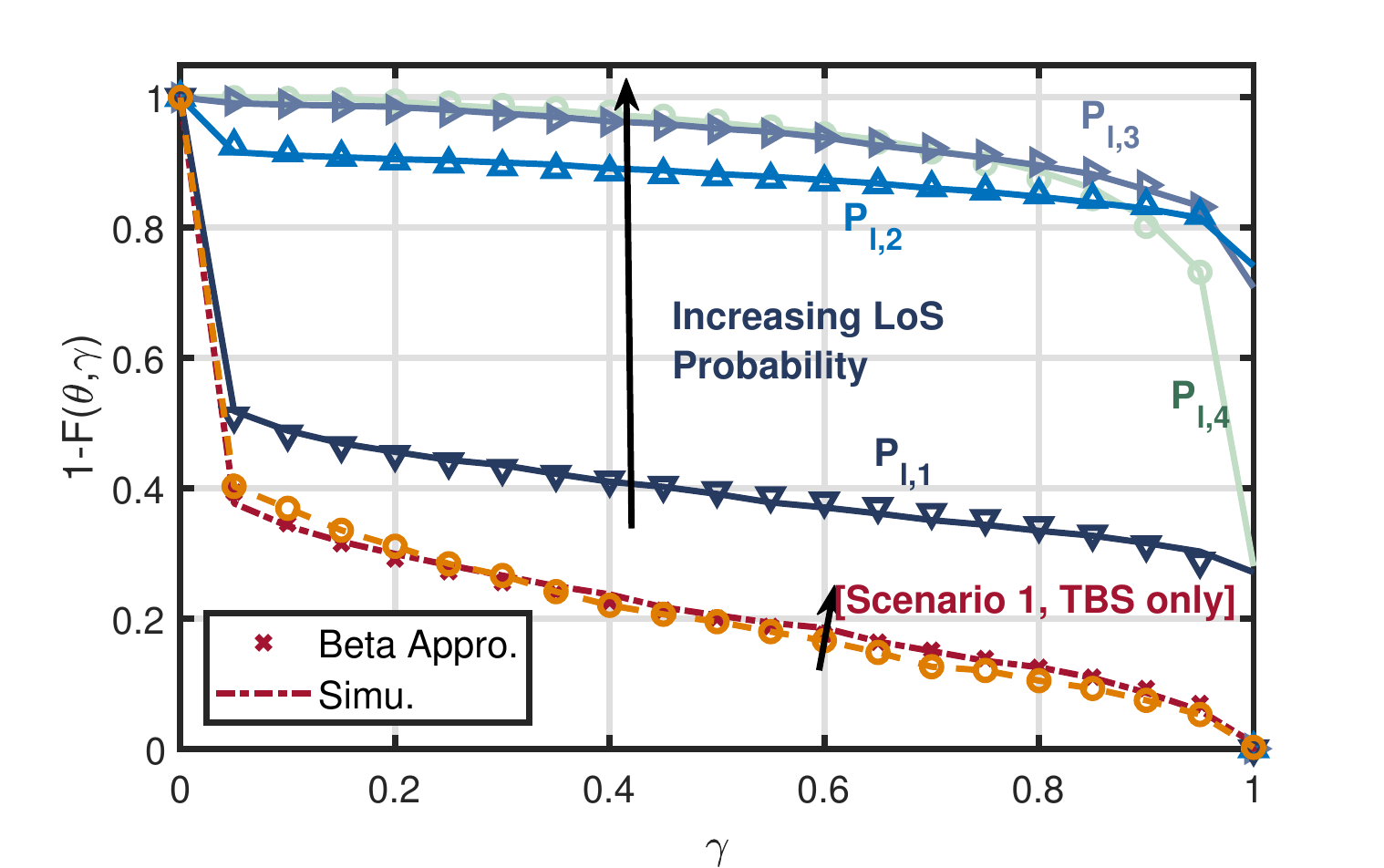}}
	\subfigure[$\theta = 10$ dB.]{\includegraphics[width=0.5\columnwidth]{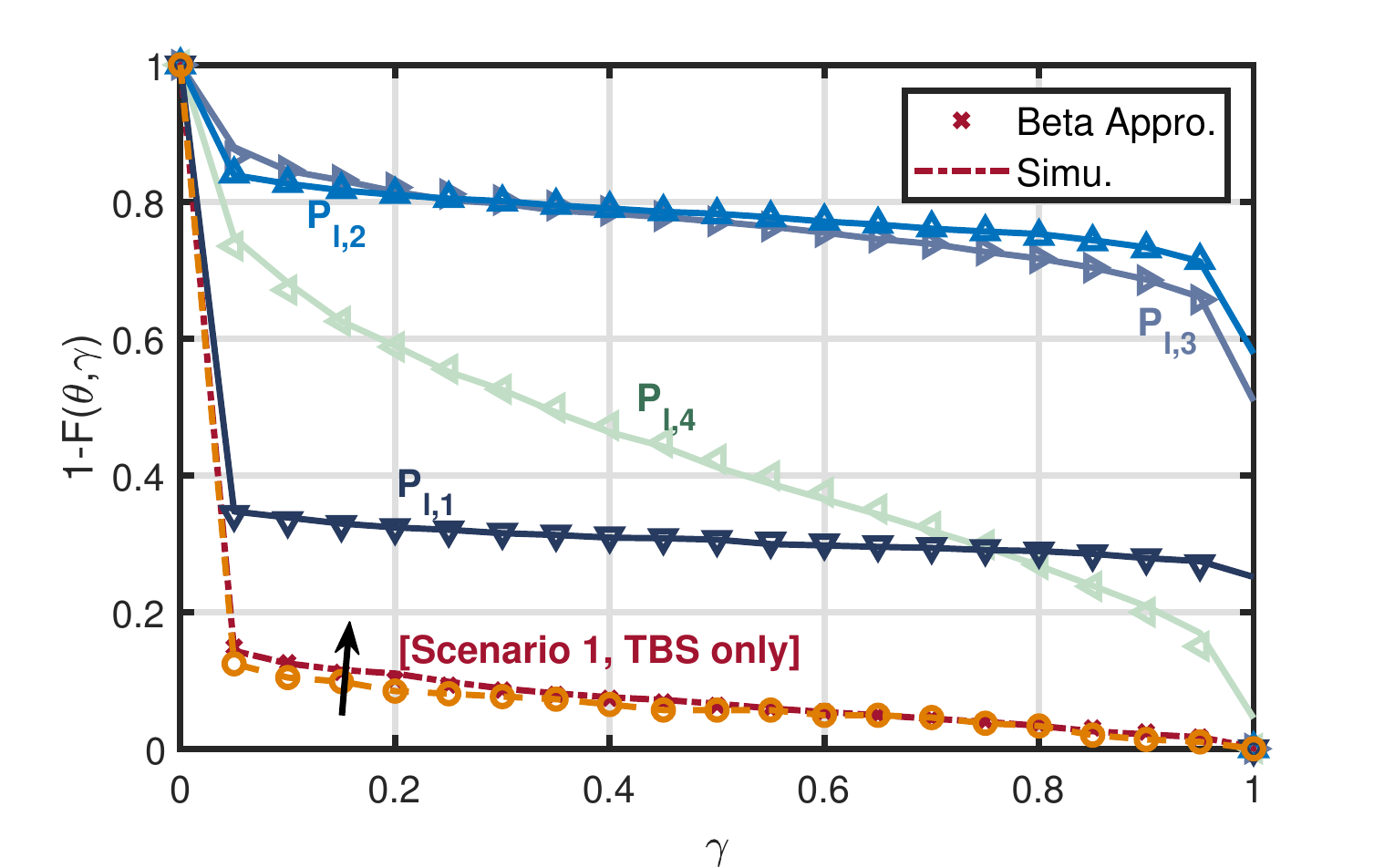}}
	\subfigure[$\theta = 20$ dB.]{\includegraphics[width=0.5\columnwidth]{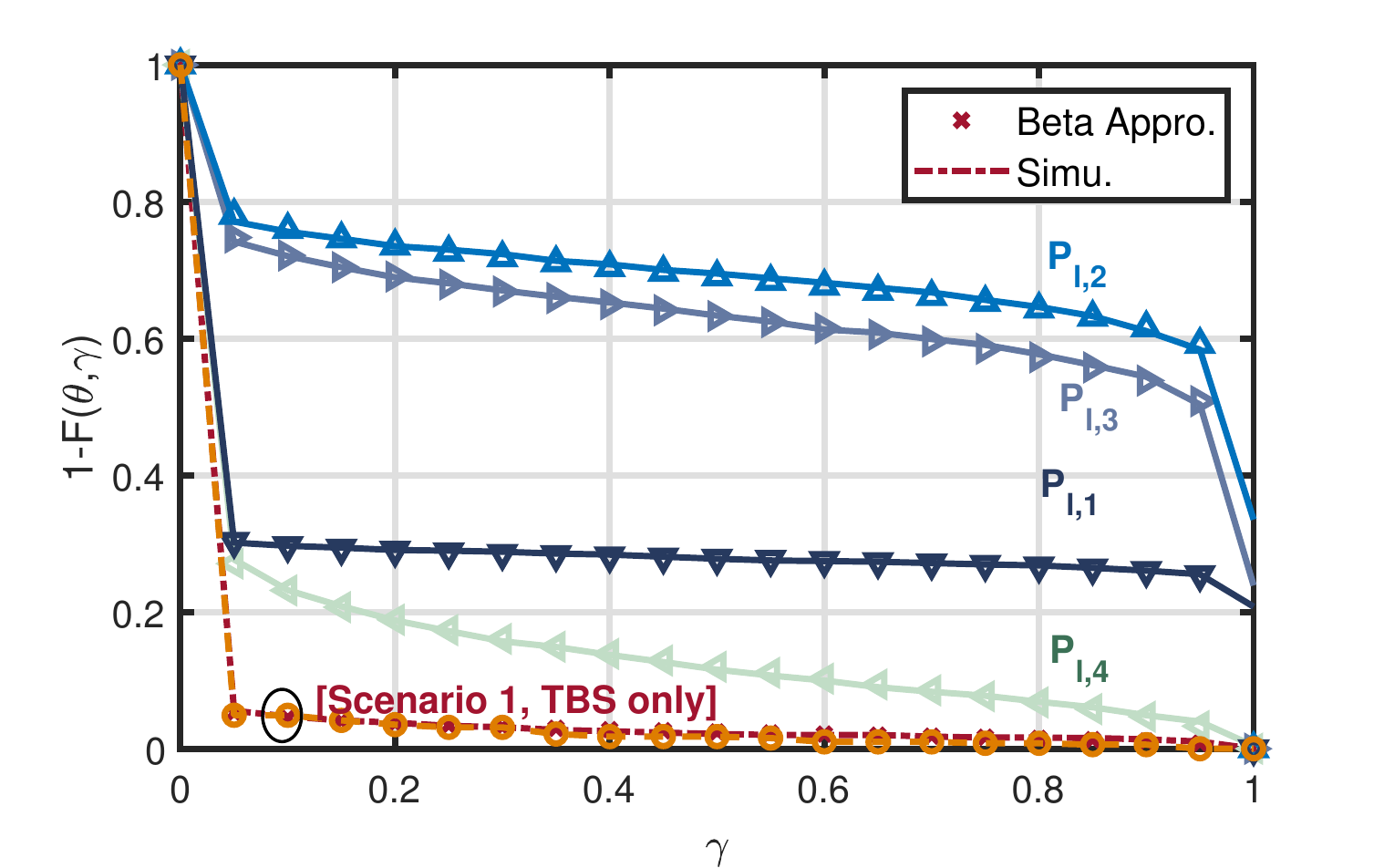}}
	\caption{The simulation and beta approximation results of the meta distribution  in the case that users are spatially-clustered, under different values of $\theta$.}
\label{Fig_4envi_difftheta}
\end{figure}

Since the SINR meta distribution is computed based on the first and the second moment, we first show the accuracy of the derived equations of the first two moments. Fig. \ref{Fig_meanvar} shows the downlink mean conditional success probability, which is also known as coverage probability, $M_1(\theta) = \mathbb{E}[P_s(\theta)]$ and the variance of conditional success probability ${\rm Var}(\theta) = M_2(\theta)-M_1^2(\theta)$ as a function of $\theta$ for TBS-only and UAV-assisted cellular networks under four different types of environments and two types of user distributions. 
The differences between traditional coverage probability ($M_1(\theta)$ in Fig. \ref{Fig_meanvar}) and SINR meta distribution can be observed by comparing Fig. \ref{Fig_meanvar} and Fig.  \ref{Fig_4envi_difftheta}. For instance, for a given $\theta$, while $M_1(\theta)$ is a constant, $\bar{F}_{P_s}(\theta)$ is a function of $\gamma$. That is, while coverage probability is the mean value of $P_s(\theta)$, SINR meta distribution is the CCDF of $P_s(\theta)$. In the case of two networks having the same coverage probability, the operators can distinguish them by comparing their CCDF.

\begin{figure}[ht]
	\subfigure[$\gamma = 0.3$.]{\includegraphics[width=0.5\columnwidth]{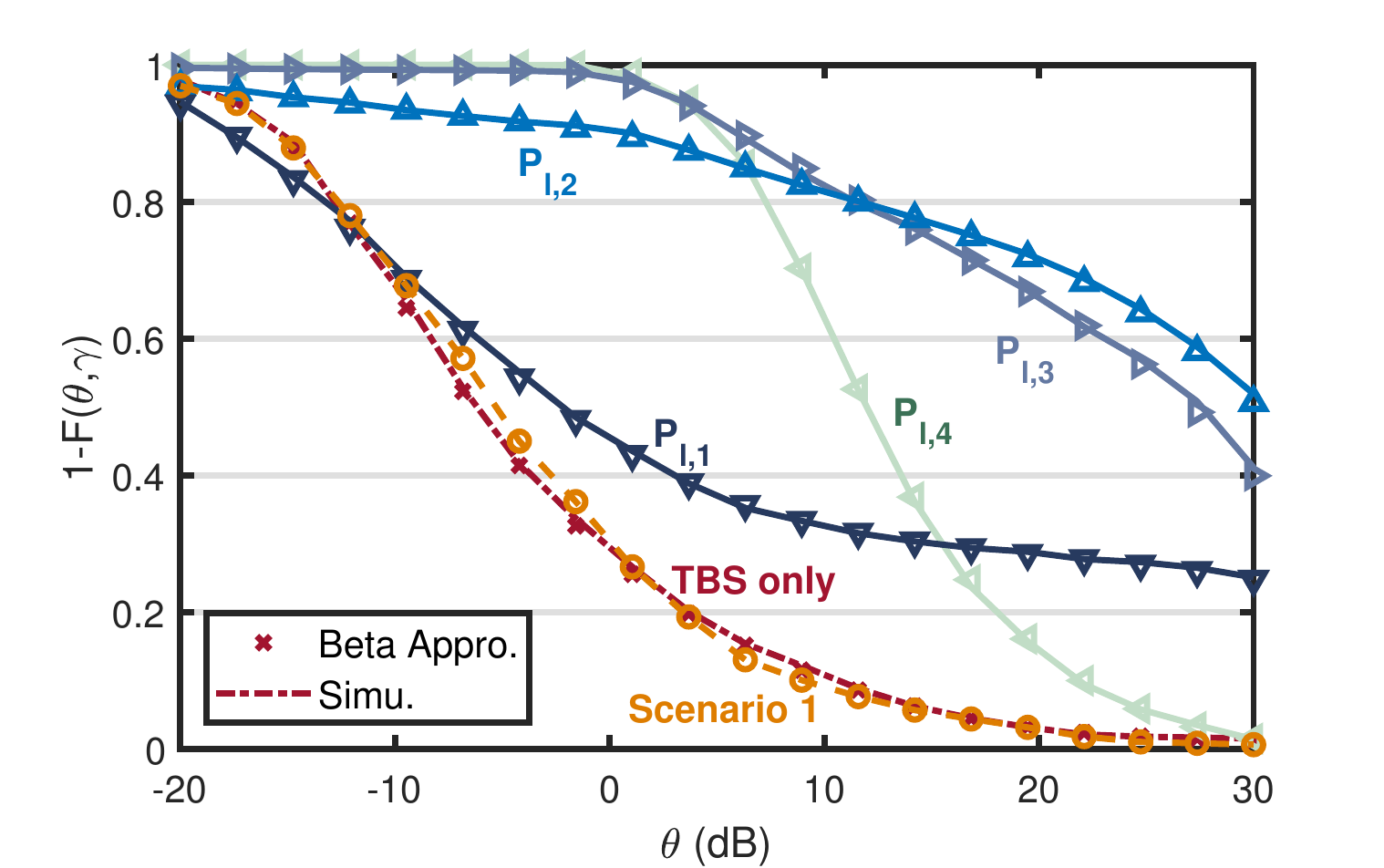}}
	\subfigure[$\gamma = 0.9$.]{\includegraphics[width=0.5\columnwidth]{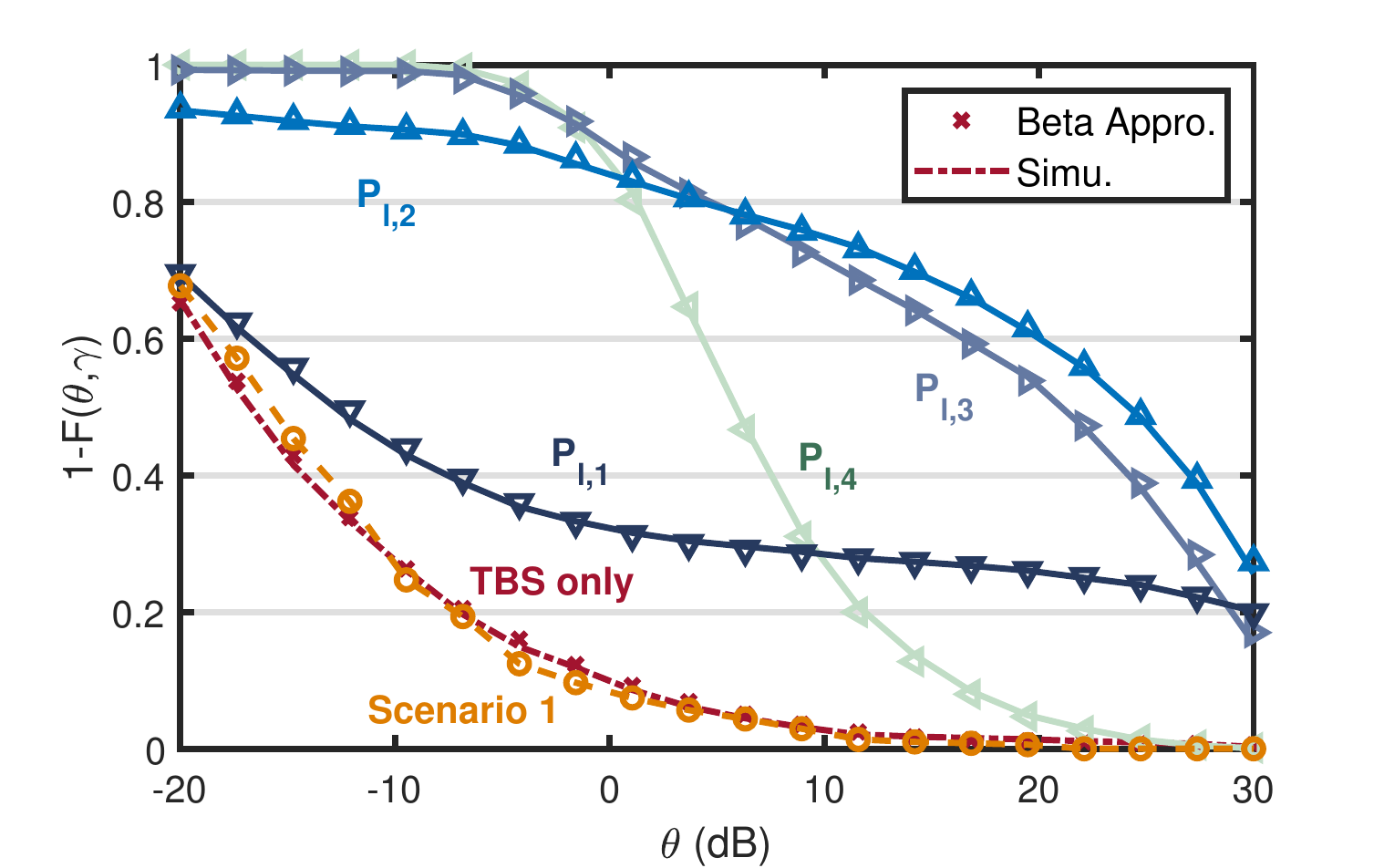}}
	\caption{The simulation and beta approximation results of the SINR meta distribution in the case that users are spatially-clustered, under different values of $\gamma$. Markers are for simulation and solid/dashed lines are for analysis.}
	\label{Fig_4envi_diffgamma}
\end{figure}

In Fig. \ref{Fig_4envi_difftheta} and Fig. \ref{Fig_4envi_diffgamma}, we investigate the impact of deploying UAVs on the reliability of the system under four environments. Fig. \ref{Fig_4envi_difftheta} shows the reliability of the system in different SINR thresholds. We also plot the results of TBS-only and scenario 1, where we convert UAVs into ground base stations. A common observation is that UAV-assisted networks are more reliable than TBS-only networks. For instance, at low SINR thresholds ($\theta < 1$) and high LoS probabilities, 80$\%$ of users can achieve a 0.8 or even higher coverage probability which is significantly greater compared with TBS-only networks, which is lower than 20$\%$ of users. Even in highrise urban areas, the percentage of users achieving a 0.8 coverage probability is almost doubled.  Besides, we observe that scenario 1 almost has no improvement on the system reliability. This is because the high-dense deploying of BSs increases the interference, and ground-to-ground communication channels have higher path loss compared to air-to-ground channels. Hence, high-dense UAV-based networks benefit from altitudes. 

Interestingly, with the increasing SINR threshold, the performance of the UAV-involved network in a high LoS probability environment drops sharply, and its performance is even worse than that in highrise urban areas. Counter-intuitively, the highest LoS probability environment case does not show the best performance since we always think LoS channels are much better than NLoS channels, and one of the advantages of UAVs is establishing LoS links with users. This can be explained by the fact that in the case of the urban or suburban areas, the interference from the LoS nearby UAVs is stronger than highrise or dense urban areas. That is, for a specific user, the probability of having LoS link with the cluster UAV increases in low-dense regions. However, the probability of having LoS interfering UAVs also increases which highly impacts the system performance.

It is worth observing that even though the deployment of UAVs in a highrise environment ($P_{l,1}(e_1 = 27,e_2=0.08)$) does not improve the system reliability a lot  compared with the other three environments in the case of low values of $\theta$, the performance is less sensitive to the SINR threshold. Its CCDF almost keeps a constant value, slightly lower than the average probability of establishing LoS link with the cluster UAV. This is because of LoS probability. As mentioned, the signal from LoS links is much better than that from NLoS links. Once the reference user and UAV establish a LoS link, it can achieve a very high SINR because of the low interference from nearby UAVs and TBSs. 

\begin{figure}
	\subfigure[Suburban areas ($e_1 = 4.88,e_2 = 0.43$), $\theta = 1$.]{\includegraphics[width=0.5\columnwidth]{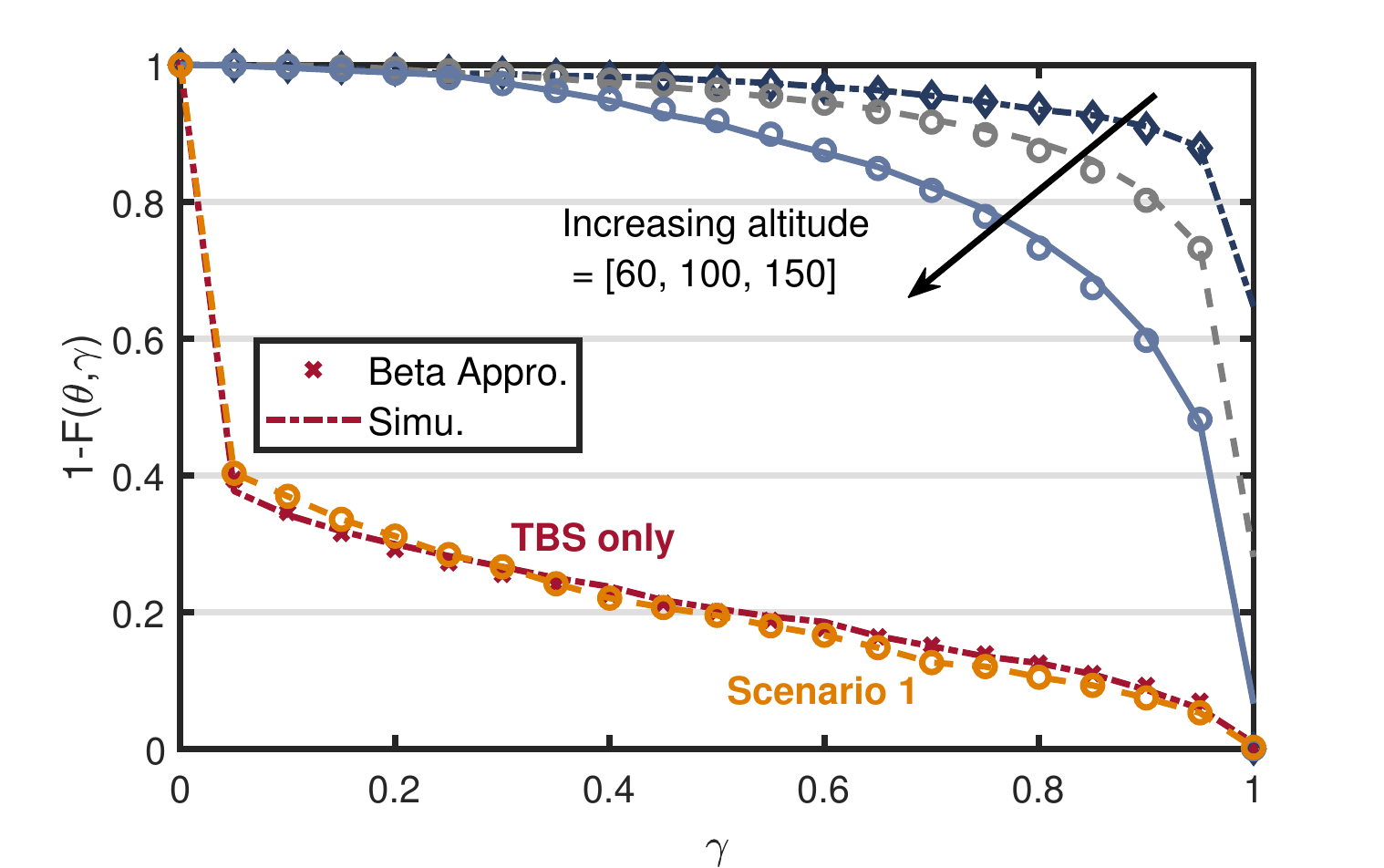}}
	\subfigure[Suburban areas ($e_1 = 4.88,e_2 = 0.43$), $\gamma = 0.9$.]{\includegraphics[width=0.5\columnwidth]{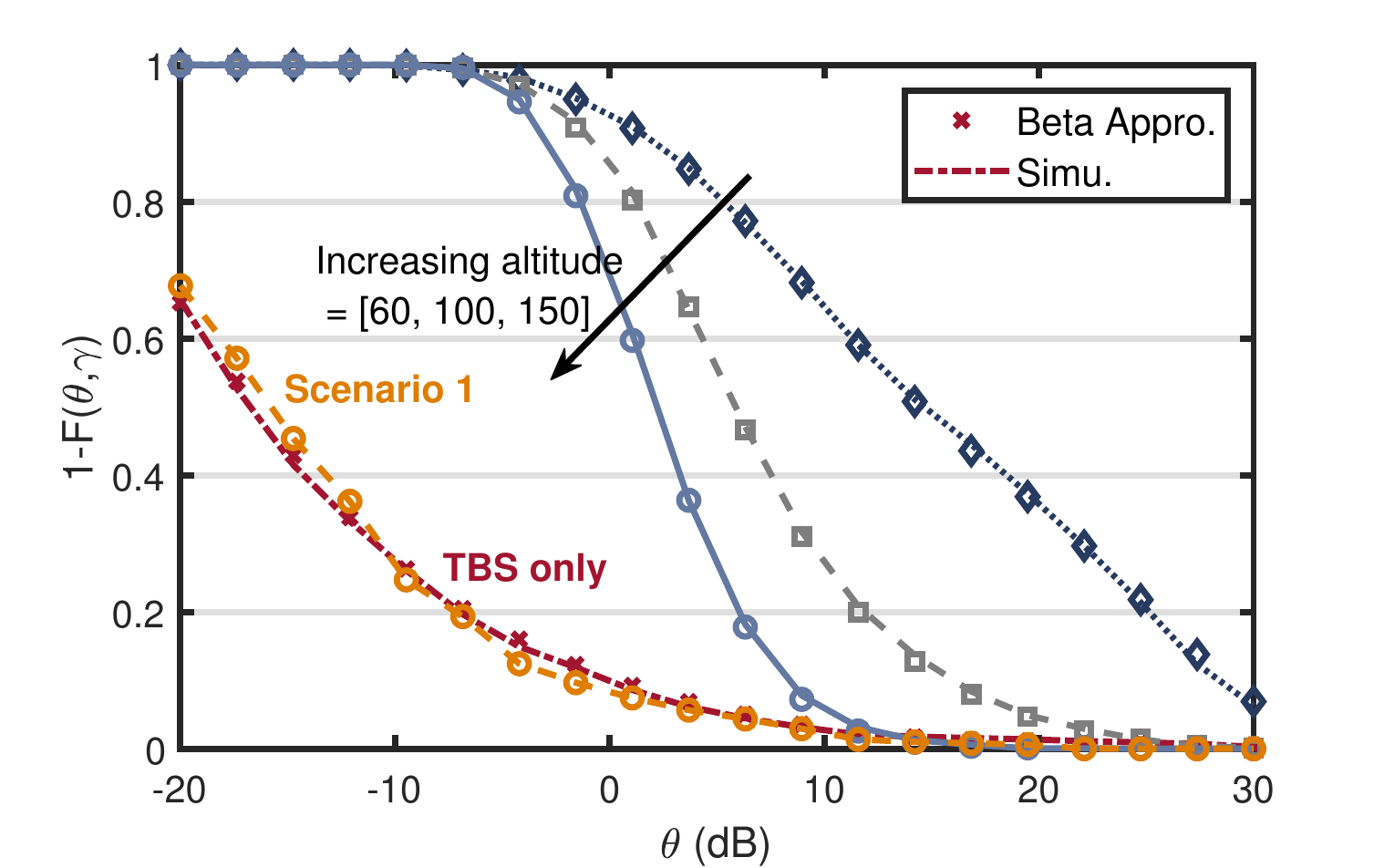}}
	\caption{The simulation and beta approximation results of the SINR meta distribution in the case of users are cluster distributed, under different values of $\theta$, under different values of the altitudes of UAVs, in the case of suburban areas. Markers are for simulation and solid/dashed lines are for analysis.}
	\label{Fig_27008_difh}
\end{figure}

 Fig. \ref{Fig_4envi_diffgamma} shows the reliability of the system in different values $\gamma$. Similar conclusions as mentioned above, the deployment of UAVs can dramatically improve the reliability of the system, while the performance of low LoS probability environment drops quickly, the performance of high-dense case is not very sensitive to the increasing of SINR threshold. We notice that for low values of $\gamma$ and $\theta$, the performance of deploying UAVs does not improve a lot or even worse, owing to deploying UAVs increasing the interference. With that being said, if the operators only require 30$\%$ of users to achieve a low SINR threshold, TBS-only networks are enough. Besides, deploying UAVs only shows improvement at very high values of $\theta$ in suburban/high dense/highrise urban areas.

\begin{figure}
\subfigure[Highrise urban areas ($e_1 = 27,e_2 = 0.08$), $\theta = 1$.]{\includegraphics[width=0.5\columnwidth]{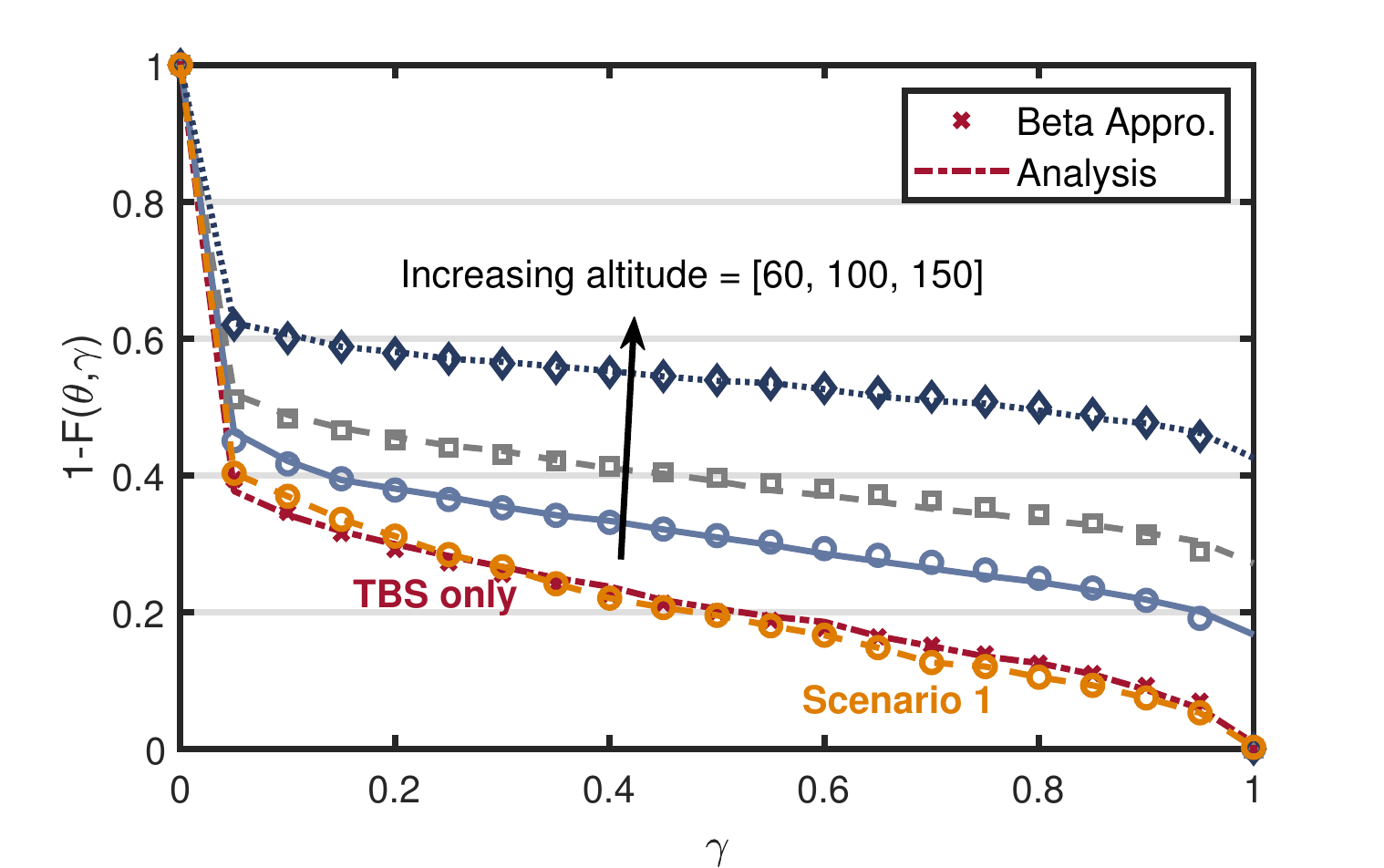}}
\subfigure[Highrise urban areas ($e_1 = 27,e_2 = 0.08$), $\gamma = 0.9$.]{\includegraphics[width=0.5\columnwidth]{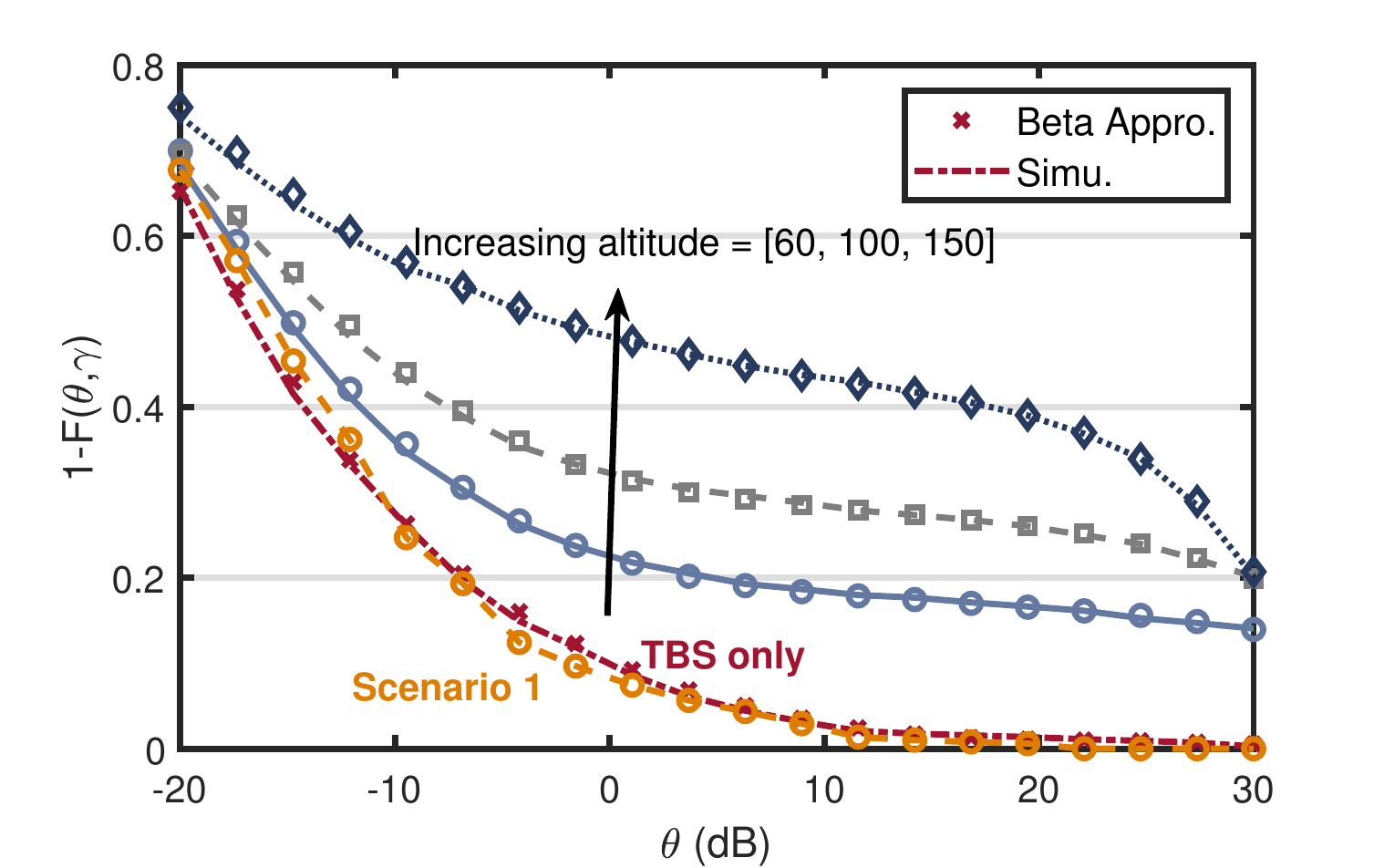}}
\caption{The simulation and beta approximation results of the SINR meta distribution of spatially-clustered users, under different values of $\theta$, under different values of the altitudes of UAVs, in the case of highrise area. Markers are for simulation and solid/dashed lines are for analysis.}
\label{Fig_488043_difh}
\end{figure}
\begin{figure}
	\centering
	\includegraphics[width=0.5\columnwidth]{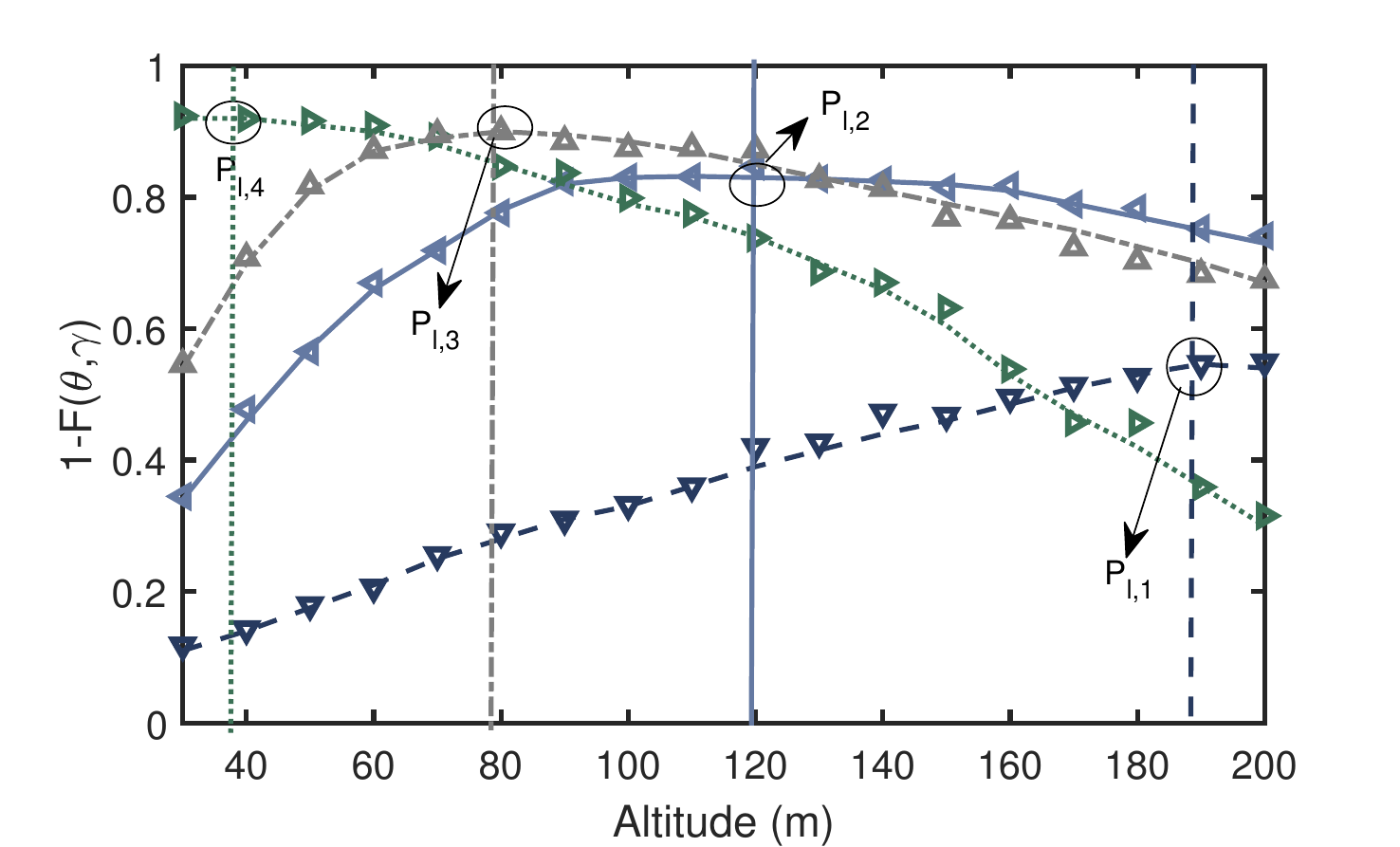}
	\caption{Optimal UAV altitudes for SINR meta distribution of spatially-clustered users under four different environments and $\theta = 0$ dB and $\gamma = 0.9$.}
	\label{Fig_OptimalAltitude}
\end{figure}

Fig. \ref{Fig_27008_difh} and Fig. \ref{Fig_488043_difh}, we plot the impact of altitudes of UAVs on the system performance. Our results reveal that for different types of environments, the optimal altitudes are different: in high LoS probability environments, we need to decrease the altitude to improve the reliability of the system, while in low LoS probability environments, we need to increase the altitudes of UAVs to increase the LoS probability, and hence, improve the system reliability. To further show the impact of UAV altitudes on the system reliability, we plot Fig. \ref{Fig_OptimalAltitude}. We show that optimal altitudes exist to maximize the system reliability and the optimal altitudes increase with the decrease of the LoS probabilities.

\begin{figure}
	\subfigure[Highrise urban areas ($e_1 = 27,e_2= 0.08$), $\gamma = 0.9$.]{\includegraphics[width=0.5\columnwidth]{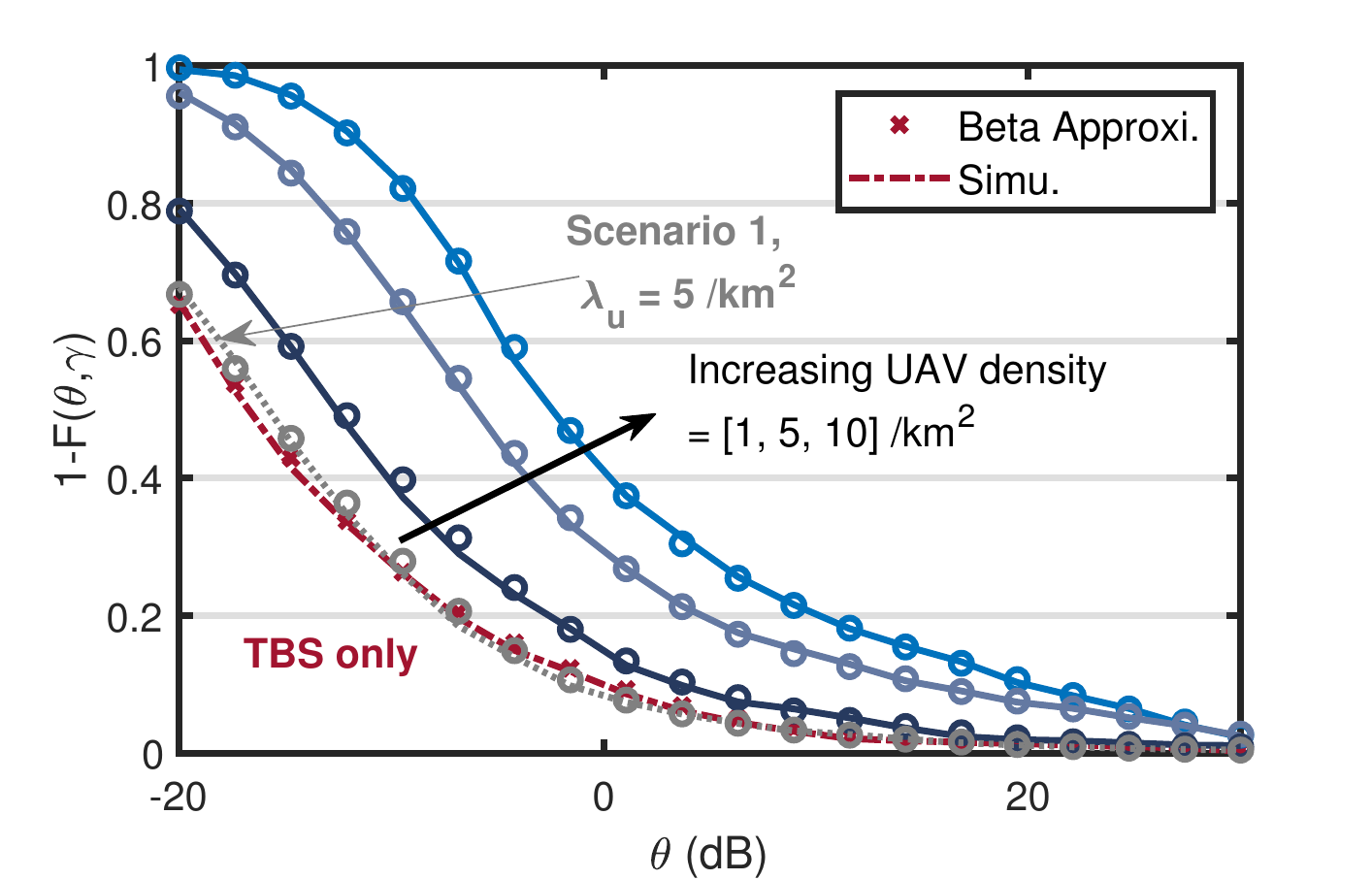}}
	\subfigure[Highrise urban areas ($e_1 = 27,e_2 = 0.08$), $\theta = 1$.]{\includegraphics[width=0.5\columnwidth]{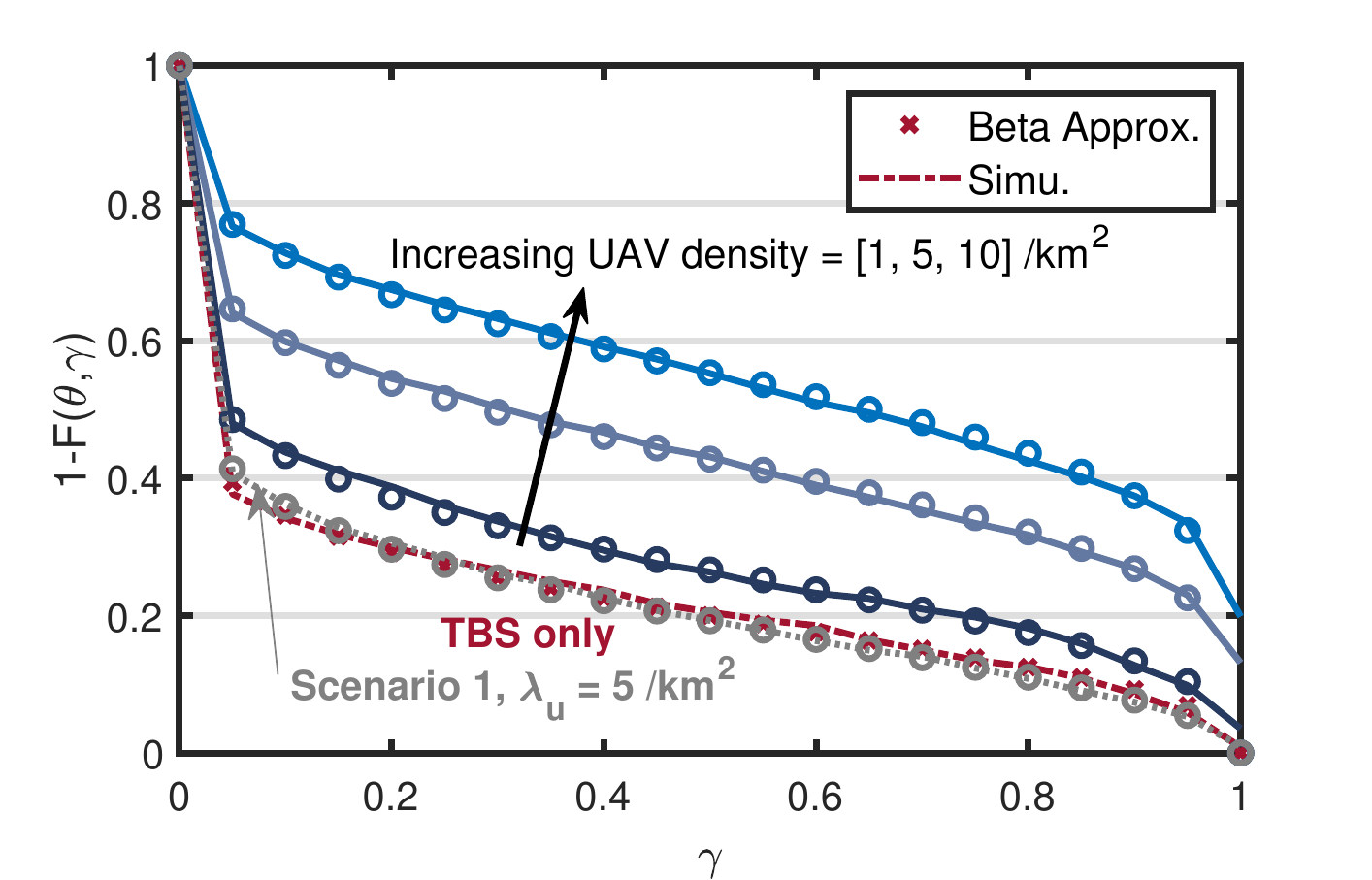}}
	\caption{The simulation and beta approximation results of the SINR meta distributions of PPP users, in the case of highrise urban areas. Markers are for simulation and solid/dashed lines are for analysis.}
	\label{Fig_density_27008}
\end{figure}
\begin{figure}
	\subfigure[Suburban areas ($e_1 = 4.88,e_2 = 0.43$), $\gamma = 0.9$.]{\includegraphics[width=0.5\columnwidth]{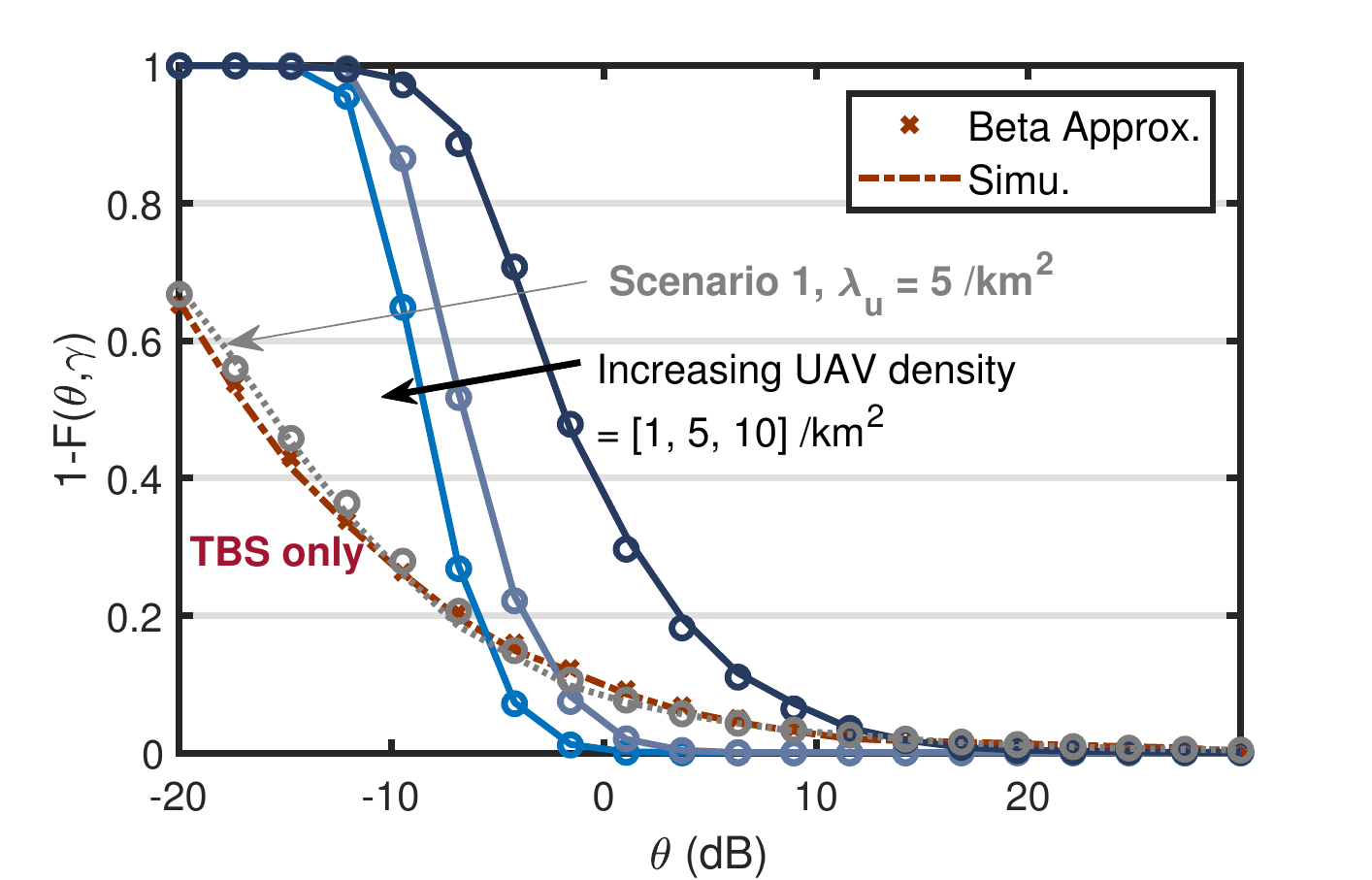}}
	\subfigure[Suburban areas ($e_1 = 4.88,e_2 = 0.43$), $\theta = 1$.]{\includegraphics[width=0.5\columnwidth]{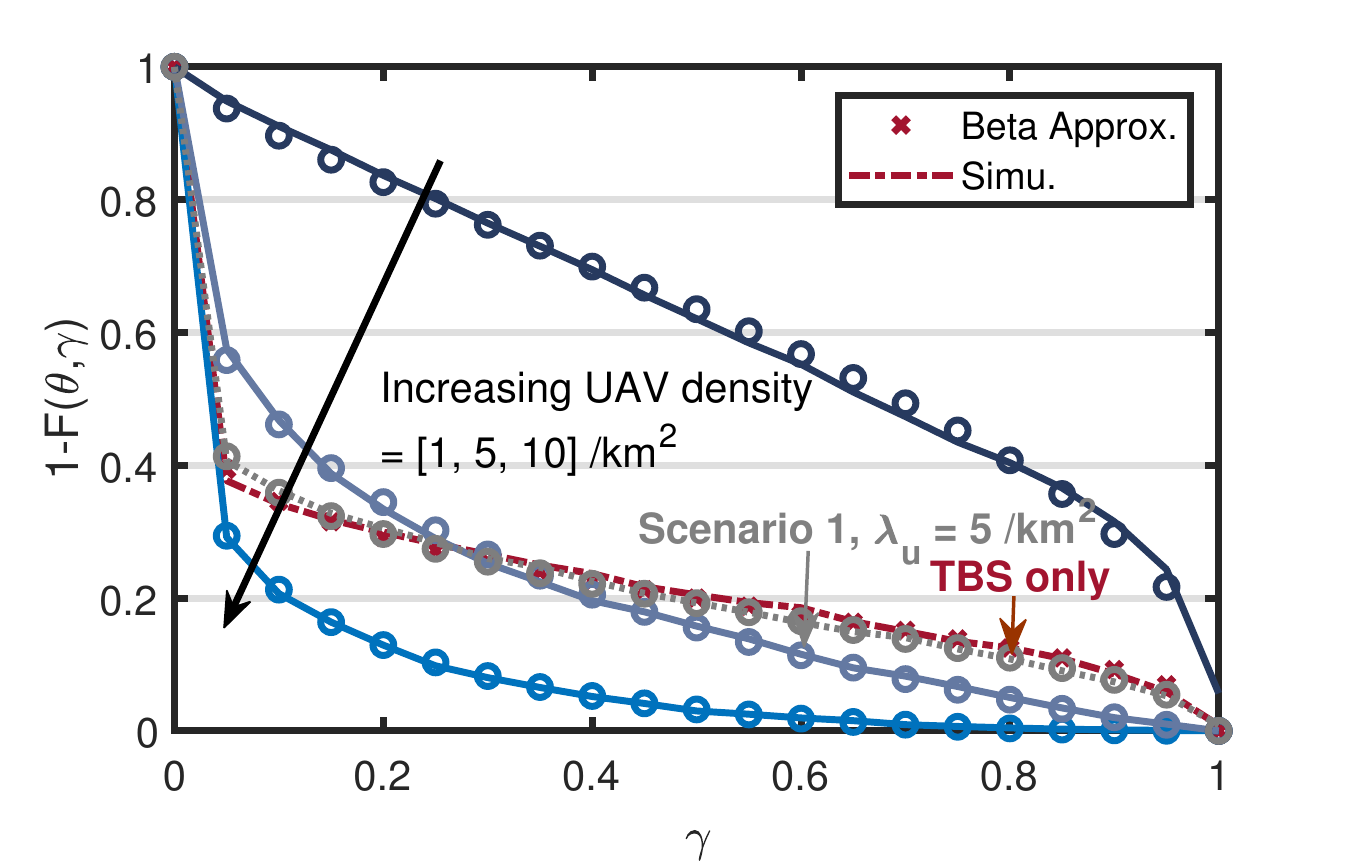}}
	\caption{The simulation and beta approximation results of the SINR meta distributions of  PPP users, in the case of suburban areas. Markers are for simulation and solid/dashed lines are for analysis.}\label{Fig_density_488043}
\end{figure}

We then plot the impact of densities of UAVs on the meta distributions in the case that users are PPP distributed in Fig. \ref{Fig_density_27008} and Fig. \ref{Fig_density_488043}. To show the performance improvement of deploying UAVs in the networks, we also plot TBS-only networks and scenario 1 (convert UAVs into ground BSs) and we only plot the curve when $\lambda_u = 5$ /km$^2$ since these three densities show similar results. We observe that scenario 1 has almost the same performance as TBS-only networks. While the high dense deployment of BSs decreases the communication distances, the interference increases.  In the case of highrise urban areas, the reliability of the network increases with the increase of the UAV density, owing to the decrease of the distances between the reference users and UAVs and the increasing probability of establishing LoS links. However, in urban areas, the network shows opposite properties. The network reliability decreases with the increasing UAV densities due to the closer LoS interfering UAVs. In addition, at high values of $\theta$, the TBS-only network achieves a better system performance.
\begin{figure}
	\centering
	\includegraphics[width=0.5\columnwidth]{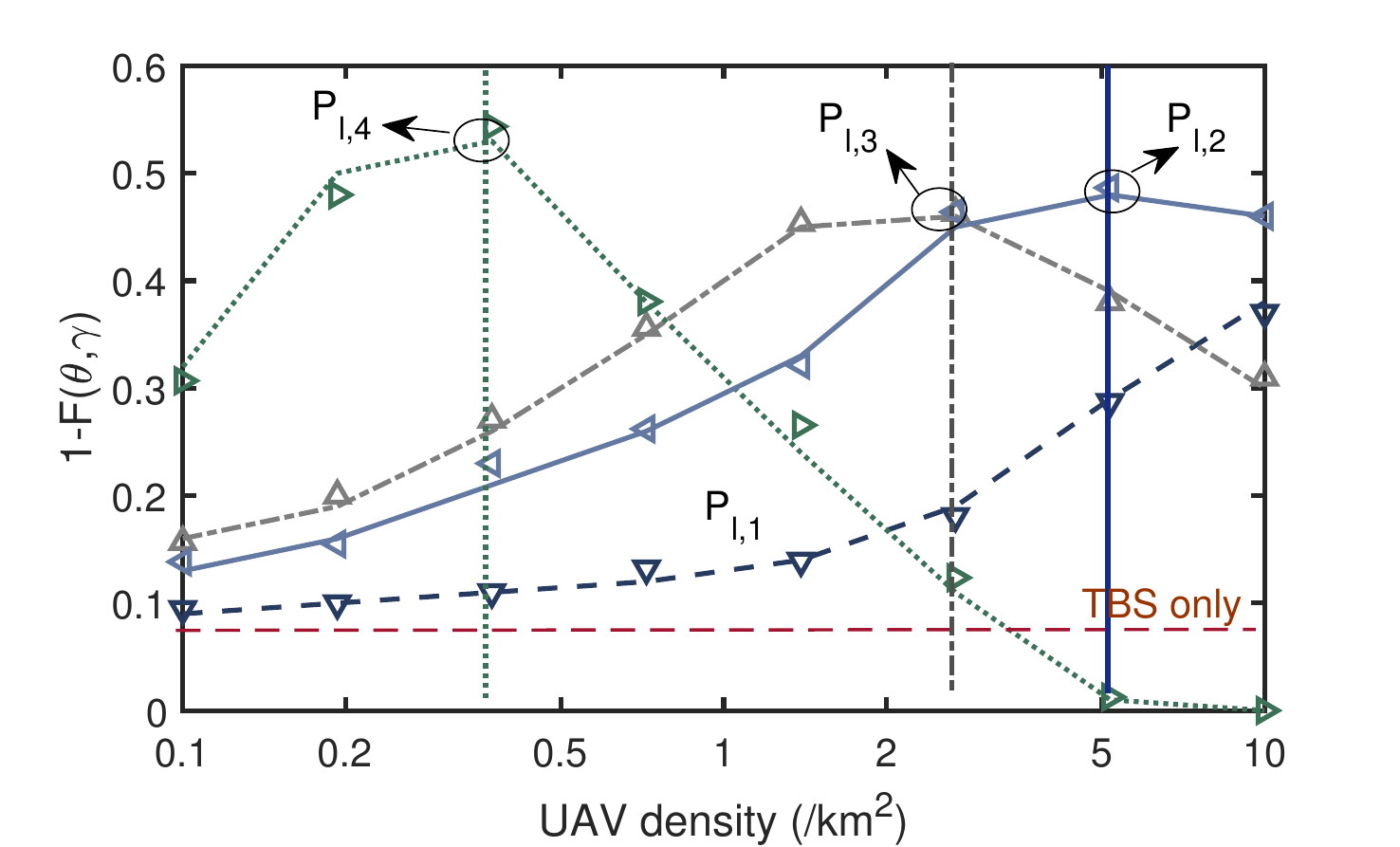}
	\caption{Optimal UAV densities for SINR meta distribution of PPP distributed users under four different environments and $\theta = 0$ dB and $\gamma = 0.9$.}
	\label{Fig_OptimalDensity}
\end{figure}
To further study the impact of UAV density on the system reliability, in Fig. \ref{Fig_OptimalDensity} we plot the SINR meta distribution under four different environments and $\theta = 0$ dB and $\gamma = 0.9$ and PPP distributed users. We show that optimal densities exist to maximize the system reliability. Optimal densities increase with the decrease of the LoS probability and the ratios of optimal UAV density to TBS density are $0.4$, $2.5$ and $5$ for suburan, urban and dense urban areas.

\begin{figure}
	\subfigure[Highrise urban areas ($e_1 = 27,e_2 = 0.08$), $\gamma = 0.9$.]{\includegraphics[width=0.5\columnwidth]{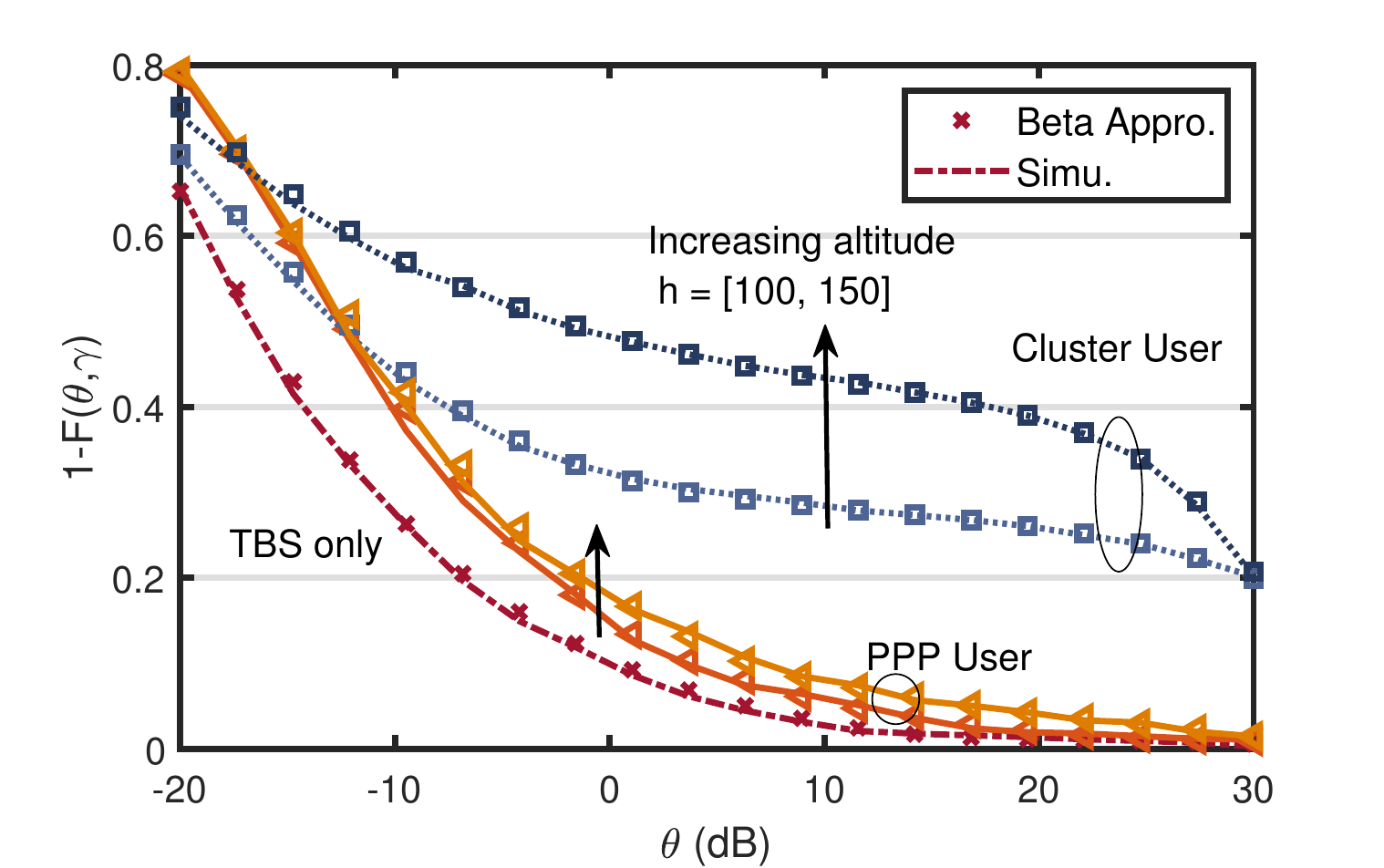}}
	\subfigure[Highrise urban areas ($e_1 = 27,e_2 = 0.08$), $\theta = 1$.]{\includegraphics[width=0.5\columnwidth]{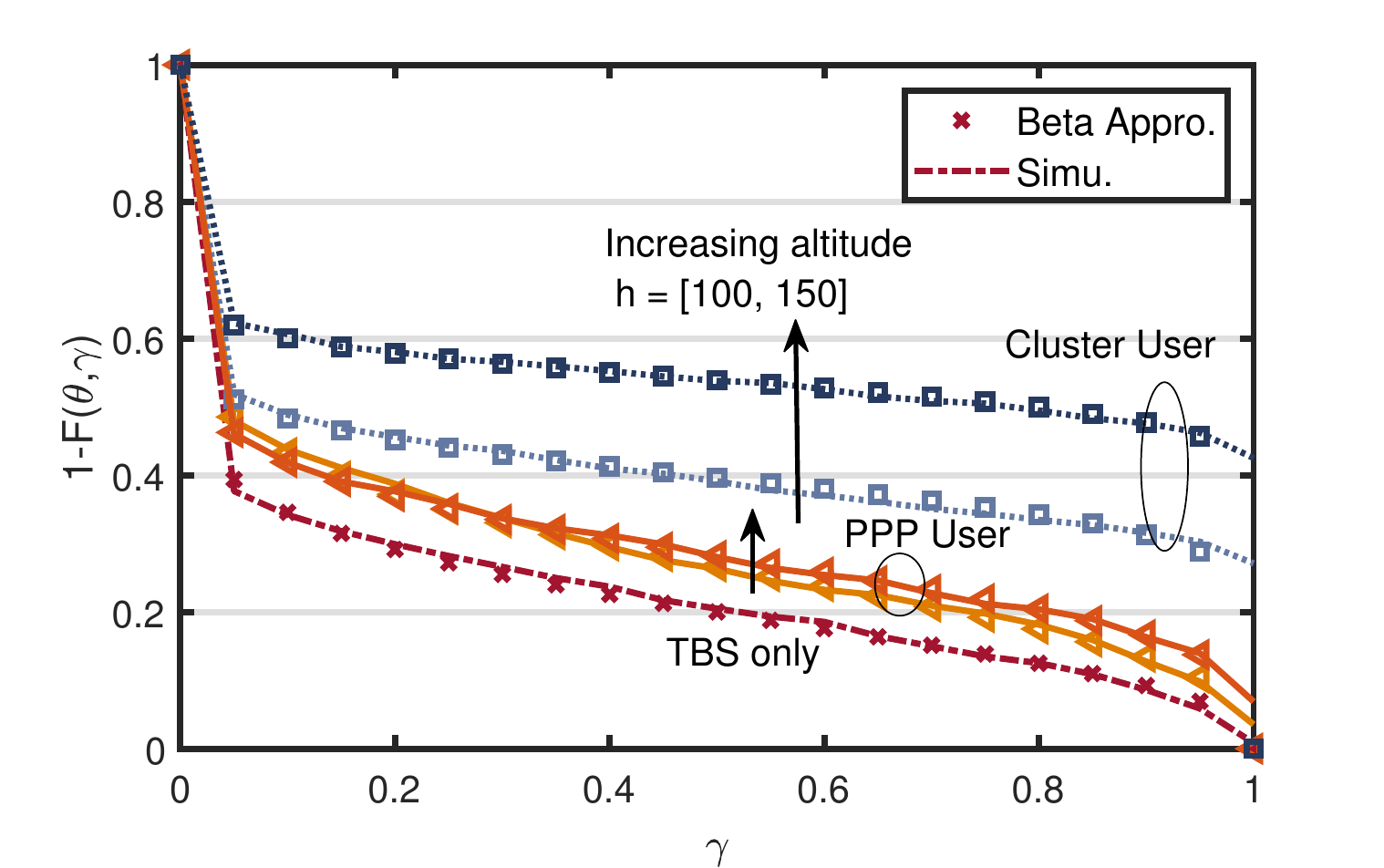}}
	\caption{The simulation and beta approximation results of the SINR meta distributions in two user distributions, PPP and cluster in the case of highrise urban areas. Markers are for simulation and solid/dashed lines are for analysis.}
	\label{Fig_compare_27008}
\end{figure}
\begin{figure}
	\subfigure[Suburban areas ($e_1 = 4.88,e_2 = 0.43$), $\gamma = 0.9$.]{\includegraphics[width=0.5\columnwidth]{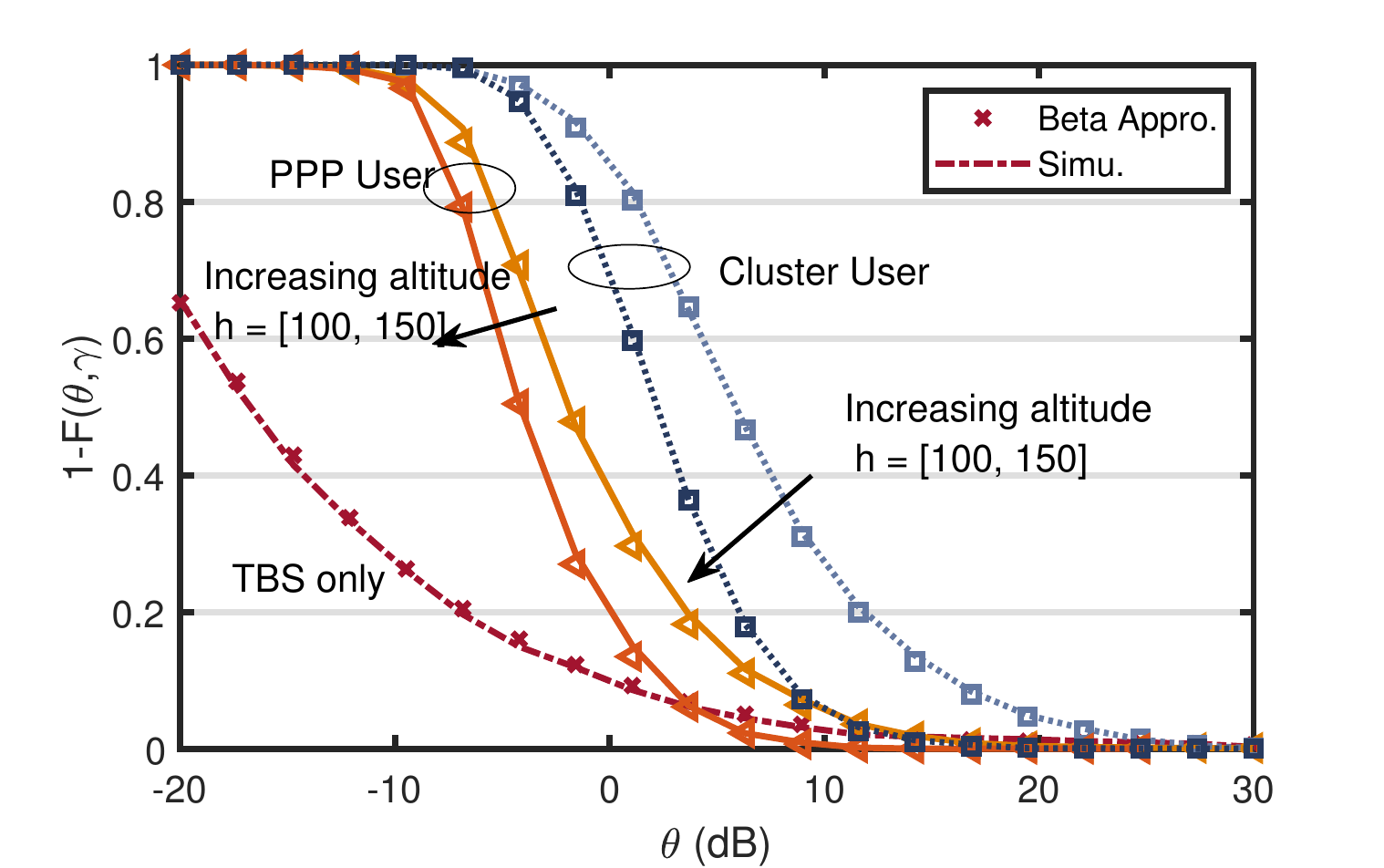}}
	\subfigure[Suburban areas ($e_1 = 4.88,e_2 = 0.43$), $\theta = 1$.]{\includegraphics[width=0.5\columnwidth]{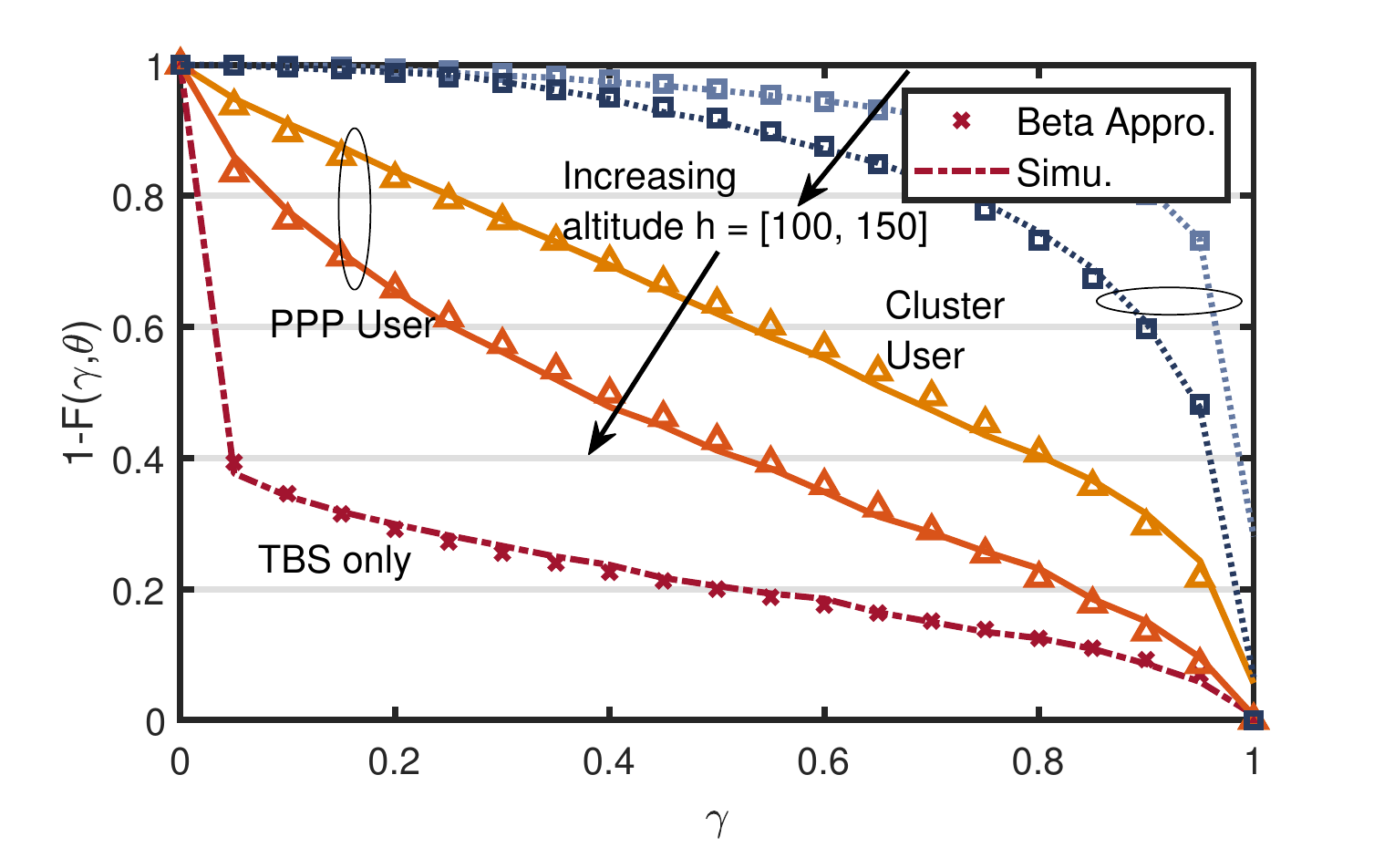}}
	\caption{The simulation and beta approximation results of the SINR meta distributions in two user distributions, PPP and cluster in the case of suburban areas. Markers are for simulation and solid/dashed lines are for analysis.}\label{Fig_compare_488043}
\end{figure}

In Fig. \ref{Fig_compare_27008} and  \ref{Fig_compare_488043}, we plot the meta distributions of two types of user distributions, PPP or MCP.  For the results of scenario 1, we omit them for both user distributions since the curves are similar to TBS-only networks. In the case of highrise urban  areas, the deployment of UAVs improves the system performance in both user distributions, while the improvement in the user cluster model is much greater. It is reasonable since deploying UAVs for user clusters fully exhibit their advantages: change their locations and deliver additional coverage to a certain area based on the dynamic demands of users. Notice that increasing the altitudes of UAVs in the case that PPP users has a lower impact on the system performance in highrise urban areas since the distances between the users and UAVs are far and LoS link probability drops quickly with the increase of distances.

\begin{figure}
	\subfigure[]{\includegraphics[width=0.5\columnwidth]{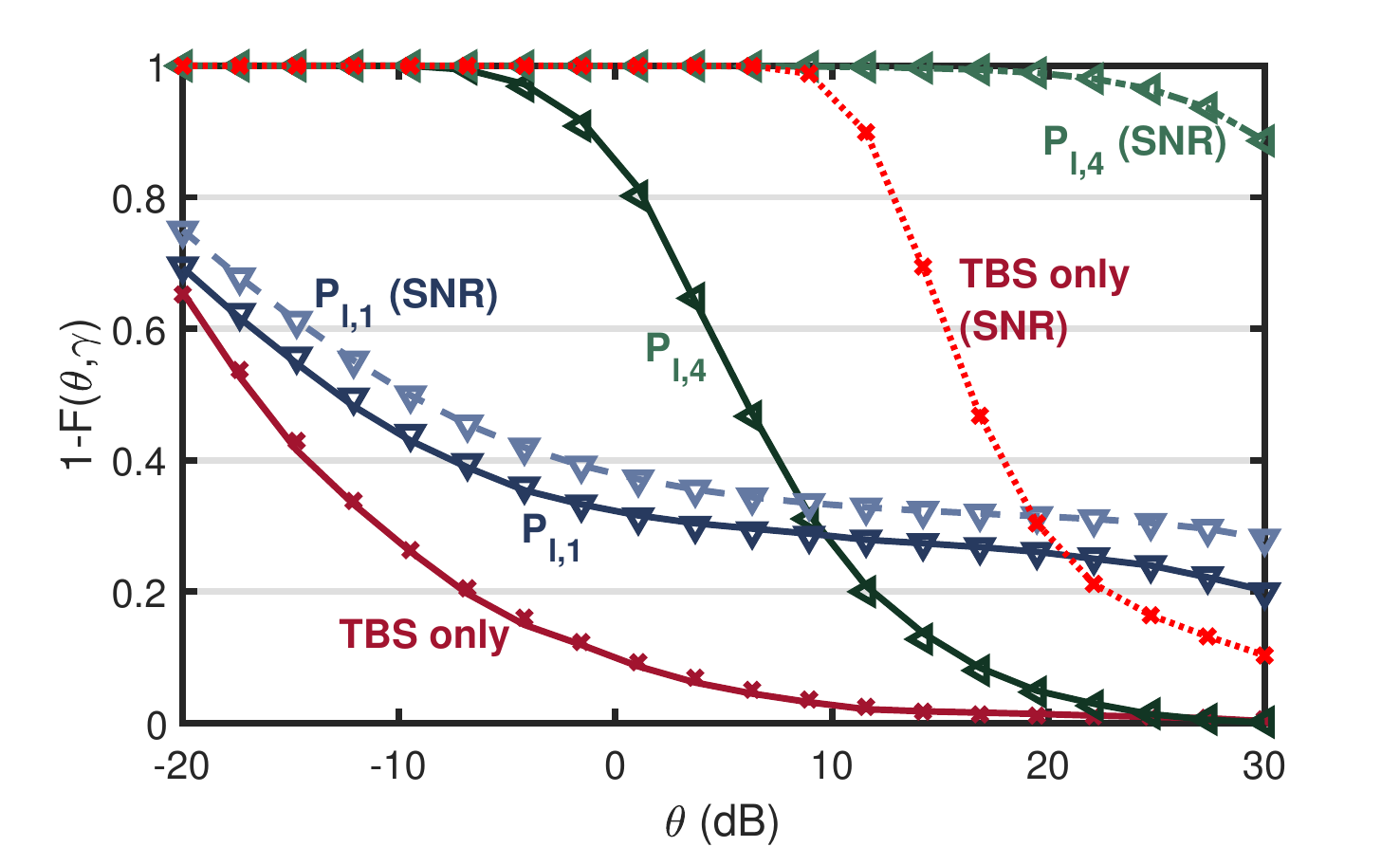}}
	\subfigure[]{\includegraphics[width=0.5\columnwidth]{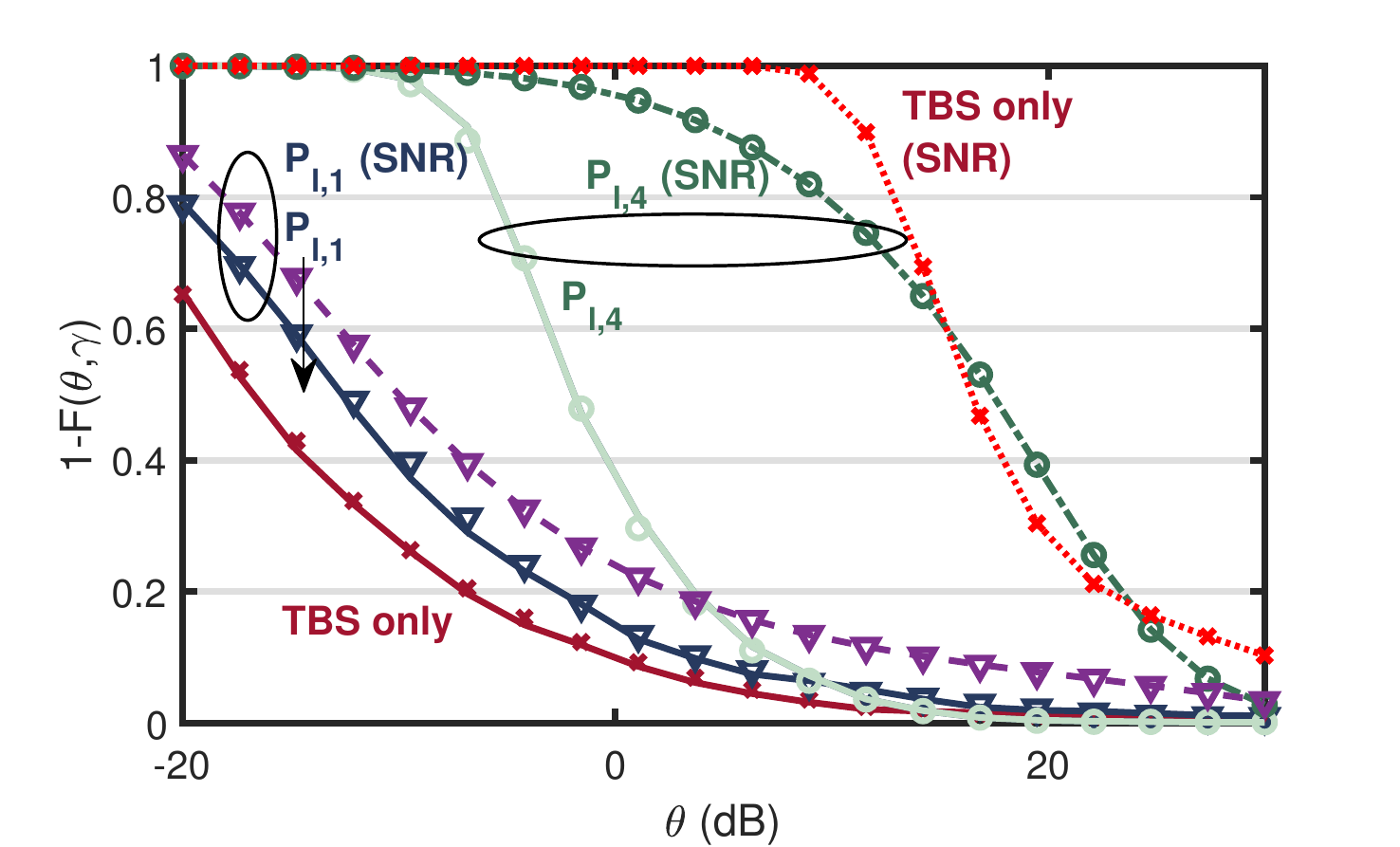}}
	\caption{The simulation and beta approximation results of the SINR (lines) and SNR (dash/dots curves) meta distributions in two user distributions, \textbf{(a)} MCP and \textbf{(b)} PPP, under suburban areas and highrise urban areas and $\lambda_{ u,\{1,2\}} = 1$/km$^2$ and $h = 100$ m.}\label{Fig_Cor}
\end{figure}
In Fig. \ref{Fig_Cor}, we plot the SNR meta distribution, which are given in  Cor. \ref{Cor_MCP}, Cor. \ref{Cor_PPP} and Cor. \ref{Cor_TBS}. We show that in highrise urban areas, SNR meta distribution is close to SINR meta distribution since interference are mostly NLoS and has limited influence on the system performance, while SNR meta distribution are quite different with SINR distribution in suburban areas. 

\section{Conclusion}
By using tools from stochastic geometry, this work presented a framework for analyzing the reliability of UAV-assisted networks. We derived analytical expressions for meta distribution of cellular networks comprising UAVs and TBSs and theoretically proved that deploying UAVs in most cases highly improves the network's reliability. We provided both the simulation and analysis results of meta distribution in four different types of environments (highrise, high dense, suburban and urban areas) and showed the impact of altitudes and densities of UAVs on the system performance. 

Our numerical results reveal several system insights. When the locations of users exhibit a certain degree of clustering, deploying UAVs can highly improve communication reliability, especially at low values of the SINR threshold. In the case of highrise urban areas, deploying a UAV for each cluster seems to only slightly improve the reliability. 
Besides, we showed that the system does not always benefit from increasing LoS probabilities: on the one side, it improves the signal. On the other side, it increases the interference and, therefore, decreases the probability of establishing extremely reliable channels (which can achieve a very high SINR, as mentioned above). For the urban area, the operators need to decrease UAVs' altitude and density to maintain a reliable network.

In addition, deploying UAVs may not be a good choice if the users are not clustered since it does not improve performance while costing a lot, especially in the high LoS probability environments or the SINR threshold required networks.

\appendix
\subsection{Proof of Lemma \ref{lemma_successprob_1}}\label{app_proof_Ps}
	The equations (\ref{eq_ps_l}), (\ref{eq_ps_n}), (\ref{eq_ps_tl}) and (\ref{eq_ps_tn}) are derived by
\begin{align}
	\mathbb{P}\bigg(\frac{p_{u,l}}{I_{u,l}+\sigma^2}>\theta\bigg)&=\bigg(G_{l}>\frac{\theta}{\eta_l \rho_u}R_{u}^{\alpha_l}(I_{u,l}+\sigma^2)\bigg)\stackrel{(a)}{=} \frac{\Gamma_{u}(m_l,m_l g_l(R_u)(I_{u,l}+\sigma^2))}{\Gamma(m_l)}\nonumber\\
	&\stackrel{(b)}{\leq} 1-(1-\exp(-s_l(R_u)(I_{u,l}+\sigma^2)))^{m_l}\nonumber\\
	&\stackrel{(c)}{=} \sum_{k_l=1}^{m_l}(-1)^{k_l+1}\exp(-k_ls_l(R_u)(I_{u,l}+\sigma^2)),\label{eq_ps_pul}\\
	\mathbb{P}\bigg(\frac{p_{u,n}}{I_{u,n}+\sigma^2}>\theta\bigg) &= \frac{\Gamma_{u}(m_n,m_n g_n(R_u)(I_{u,n}+\sigma^2))}{\Gamma(m_n)} \nonumber\\
	&\leq \sum_{k_n=1}^{m_n}(-1)^{k_n+1}\exp(-k_ns_n(R_u)(I_{u,n}+\sigma^2)), \label{eq_ps_pun}\\
	\mathbb{P}\bigg(\frac{p_t}{I_{\{tl,tn\}}+\sigma^2}>\theta\bigg) &= \mathbb{P}\bigg(h_0>\frac{\theta}{ \rho_t}R_{t}^{\alpha_t}(I_{\{tl,tn\}}+\sigma^2)\bigg)= \exp(-s_t(R_t)(I_{\{tl,tn\}}+\sigma^2)),\label{eq_proof_ps_t}
\end{align}
where $(a)$ follows from the definition of Gamma distribution, $(b)$ follows from the upper bound of Gamma distribution \cite{galkin2019stochastic,alzer1997some}: $\frac{\Gamma_{l}(m,mg)}{\Gamma(m)}<(1-\exp(-\beta_2(m)mg))^{m}$, where $\Gamma_{l}(m,mg)$ is the lower incomplete Gamma function \cite{bai2014coverage}, and $(c)$ results from the binomial theorem.  (\ref{eq_ps_pun}) follows the same steps as (\ref{eq_ps_pul}) and the equal sign comes from the fact that the shape and scale parameters $(m_n,\frac{1}{m_n})$ can equal to $(1,1)$. (\ref{eq_proof_ps_t}) is derived by using the CDF of exponential distribution.

\subsection{Proof of Theorem \ref{theorem_Mb_1}}\label{app_proof_Mb}
	In this part, we provide the proof of the $b$-th moment of the conditional success probability. 
\begin{align}
		M_b(\theta) &= \mathbb{E}[P_{s}^{b}(\theta)]\stackrel{(a)}{=} \mathbb{E}[P_{s,l}^b(\theta)+P_{s,n}^b(\theta)+P_{s,tl}^b(\theta)+P_{s,tn}^b(\theta)]\nonumber\\
		&= M_{b,l}(\theta)+M_{b,n}(\theta)+M_{b,tl}(\theta)+M_{b,tn}(\theta).
	\end{align}
where the step $(a)$ results from using the indicator function (more details are provided in Remark \ref{rem_bth_moment}).
	(\ref{eq_mb_l}) is derived by substituting (\ref{eq_ps_pul}) into (\ref{eq_mb}) and averaging over the distance,
\begin{align}
	M_{b,l}(\theta) &= \mathbb{E}[P_{s,l}^b(\theta)] = \nonumber\\
	&\mathbb{E}\bigg[ 	\sum_{k_1=1}^{m_l}\sum_{k_2=1}^{m_l}\cdots\sum_{k_b=1}^{m_l}\binom{m_l}{k_1}\binom{m_l}{k_2}\cdots\binom{m_l}{k_b}(-1)^{k_1+k_2+\cdots+k_b+b}\mathbbm{1}({\rm LoS})\mathbbm{1}(R_t>d_{lt}(R_u))\nonumber\\
	&\prod_{T_k\in\Phi_{ t}}\bigg(\frac{1}{1+k_1s_l(R_u)\rho_t D_{T_k}^{-\alpha_t}}\bigg)\bigg(\frac{1}{1+k_2s_l(R_u)\rho_t D_{T_k}^{-\alpha_t}}\bigg)\cdots\bigg(\frac{1}{1+k_bs_l(R_u)\rho_t D_{T_k}^{-\alpha_t}}\bigg)\nonumber\\
	&\prod_{N_i\in\Phi_{ u_n}}\bigg[\frac{m_n}{m_n+k_1s_l(R_u)\eta_n\rho_u D_{N_i}^{-\alpha_n}}\frac{m_n}{m_n+k_2s_l(R_u)\eta_n\rho_u D_{N_i}^{-\alpha_n}}\cdots\frac{m_n}{m_n+k_bs_l(R_u)\eta_n\rho_u D_{N_i}^{-\alpha_n}}\bigg]^{m_n}\nonumber\\
	&\prod_{L_j\in\Phi_{ u_l}}\bigg[\frac{m_l}{m_l+k_1s_l(R_u)\eta_l\rho_u D_{L_j}^{-\alpha_l}}\frac{m_l}{m_l+k_2s_l(R_u)\eta_l\rho_u D_{L_j}^{-\alpha_l}}\cdots\frac{m_l}{m_l+k_bs_l(R_u)\eta_l\rho_u D_{L_j}^{-\alpha_l}}\bigg]^{m_l}\nonumber\\
	&\exp(-(k_1+k_2+\cdots +k_b)s_l(R_u)\sigma^2)\bigg],
\end{align}
the proof completes by applying PGFL of inhomogeneous PPP and (\ref{eq_f}). Similarly, the $b$-th moment of $P_{s,n}(\theta)$ follows a similar method and thus omitted here.

The detailed steps of computing (\ref{eq_mb_tl}) are shown as follows, 
\begin{align}
	M_{b,tl}(\theta)
	&= \mathbb{E}\bigg[ \mathbbm{1}({\rm LoS})\mathbbm{1}(R_t<d_{lt}(R_u))\prod_{T_k\in\Phi_{ t}\setminus \{t_0\}}\bigg(\frac{1}{1+s_t(R_t)\rho_t D_{T_k}^{-\alpha_t}}\bigg)^b\bigg(\frac{m_l}{m_l+s_t(R_t)\eta_l\rho_u R_{u}^{-\alpha_l}}\bigg)^{b m_l}\nonumber\\
	&\prod_{N_i\in\Phi_{ u_n}}\bigg(\frac{m_n}{m_n+s_t(R_t)\eta_n\rho_u D_{N_i}^{-\alpha_n}}\bigg)^{b m_n}\prod_{L_j\in\Phi_{ u_l}}\bigg(\frac{m_l}{m_l+s_t(R_t)\eta_l\rho_u D_{L_j}^{-\alpha_l}}\bigg)^{b m_l}\exp(- b s_t(R_t)\sigma^2)\bigg]\nonumber\\
	&= \mathbb{E}\bigg[ P_l(\sqrt{R_u^2-h^2})\mathbbm{1}(R_t<d_{lt}(R_u))\mathcal{L}(b,s_t(R_t))\nonumber\\
	&\exp\bigg(-2\pi\lambda_t\int_{R_t}^{\infty}\bigg[1-\bigg(\frac{1}{1+s_t(R_t) \rho_t z^{-\alpha_t} }\bigg)^b\bigg]z{\rm d}z\bigg)\bigg(\frac{m_l}{m_l+s_t(R_t)\eta_l\rho_u R_{u}^{-\alpha_l}}\bigg)^{b m_l}\bigg]\nonumber\\
	&\stackrel{(a)}{=} \mathbb{E}\bigg[ \int_{0}^{d_{lt}(R_u)}2\pi x\lambda_t\exp(-\pi x^2\lambda_t) P_l(\sqrt{R_u^2-h^2}) \mathcal{L}(b,s_t(x))\nonumber\\
	&\exp\bigg(-2\pi\lambda_t\int_{x}^{\infty}\bigg[1-\bigg(\frac{1}{1+s_t(x) \rho_t z^{-\alpha_t} }\bigg)^b\bigg]z{\rm d}z\bigg)\bigg(\frac{m_l}{m_l+s_t(x)\eta_l\rho_u R_{u}^{-\alpha_l}}\bigg)^{b m_l}{\rm d}x\bigg]\nonumber\\
	&=\int_{h}^{\sqrt{r_c^2+h^2}}  \int_{0}^{d_{lt}(y)}2\pi x\lambda_t\exp(-\pi x^2\lambda_t) P_l(\sqrt{y^2-h^2}) \mathcal{L}(b,s_t(x))\nonumber\\
	&\exp\bigg(-2\pi\lambda_t\int_{x}^{\infty}\bigg[1-\bigg(\frac{1}{1+s_t(x) \rho_t z^{-\alpha_t} }\bigg)^b\bigg]z{\rm d}z\bigg)\bigg(\frac{m_l}{m_l+s_t(x)\eta_l\rho_u y^{-\alpha_l}}\bigg)^{bm_l}{\rm d}x\frac{2y}{r_c^2}{\rm d}y,
\end{align}
in which step $(a)$ results from the indicator function of $R_t$, and as mentioned above, we write the interference term into two Laplace transform to simplify the equation.  The derivation of (\ref{eq_mb_tn}) follows same steps as (\ref{eq_mb_tl}), thus omitted here.

\bibliographystyle{IEEEtran}
\bibliography{Ref8}
\end{document}